\documentclass{elsarticle}
\pdfoutput=1

\usepackage{amssymb}

\newcounter{bla}

\journal{Computer Physics Communications}

\usepackage{natbib}
\usepackage{geometry}
\usepackage{fleqn}
\usepackage{graphicx}
\usepackage[pagebackref]{hyperref} 

\usepackage{amsmath}

\usepackage{amssymb}
\usepackage{color}
\usepackage[dvipsnames]{xcolor}
\usepackage{xspace}                                                    
\usepackage{float} 
\usepackage[normalem]{ulem} 
\usepackage{enumitem}
\setitemize{itemsep=1ex,topsep=0pt,parsep=0pt,partopsep=0pt}
\usepackage{subcaption}

\newcommand{\nlox}{\textnormal{\textsc{NLOX}}\xspace}
\newcommand{\nloxutil}{\textnormal{\textsc{NLOX}\texttt{\_util}}\xspace}
\newcommand{\tred}{\textnormal{\texttt{TRed}}\xspace}
\newcommand{\polythree}{\textnormal{\texttt{Poly3}}\xspace}
\newcommand{\blackhat}{\textnormal{\textsc{BlackHat}}\xspace}

\newcommand{\gosam}{\textnormal{\textsc{GoSam}}\xspace}
\newcommand{\helaconeloop}{\textnormal{\textsc{Helac-1Loop}\xspace}}
\newcommand{\madloop}{\textnormal{\textsc{MadLoop}}\xspace}
\newcommand{\openloops}{\textnormal{\textsc{OpenLoops}}\xspace}
\newcommand{\recola}{\textnormal{\textsc{Recola}}\xspace}
\newcommand{\mgfive}{\textnormal{\textsc{MG5\_aMC@NLO}}\xspace}
\newcommand{\powhegbox}{\textnormal{\textsc{PowhegBox}}\xspace}
\newcommand{\sherpa}{\textnormal{\textsc{Sherpa}}\xspace}
\newcommand{\herwigseven}{\textnormal{\textsc{Herwig7}}\xspace}

\newcommand{\vbfnlo}{\textnormal{\textsc{VBFNLO}}\xspace}
\newcommand{\njet}{\textnormal{\textsc{NJet}}\xspace}
\newcommand{\mcfm}{\textnormal{\textsc{MCFM}}\xspace}
\newcommand{\oneloop}{\textnormal{\textsc{OneLOop}}\xspace}
\newcommand{\qcdloop}{\textnormal{\textsc{QCDLoop}}\xspace}
\newcommand{\looptools}{\textnormal{\textsc{LoopTools}}\xspace}
\newcommand{\python}{\textnormal{\texttt{Python}}\xspace}
\newcommand{\cpp}{\textnormal{\texttt{C++}}\xspace}

\newcommand{\fortran}{\textnormal{\texttt{Fortran}}\xspace}
\newcommand{\gcc}{\textnormal{\texttt{gcc}}\xspace}
\newcommand{\qgraf}{\textnormal{\textsc{QGRAF}}\xspace}
\newcommand{\form}{\textnormal{\textsc{FORM}}\xspace}
\newcommand{\sla}{\!\!\!/}
\newcommand{\order}[1]{\mathcal{O}(#1)}
\newcommand{\eps}{\epsilon}

\newcommand{\eg}{\textit{e.g.}\xspace} 
\newcommand{\ie}{\textit{i.e.}\xspace}

\newcommand{\lo}{lowest-order\xspace}
\newcommand{\nlo}{next-to-lowest-order\xspace}
\newcommand{\ir}{\mathrm{ir}}
\newcommand{\uv}{\mathrm{uv}}

\newcommand{\msbar}{\overline{\mathrm{MS}}}
\newcommand{\amp}{A}

\begin{document}

\begin{frontmatter}



\title{\nlox, a one-loop provider for Standard Model processes}


\author[a]{Steve Honeywell\fnref{SH}}
\author[a]{Seth Quackenbush\fnref{SQ}}
\author[a]{Laura Reina\fnref{LR}}
\author[a,b]{Christian Reuschle\fnref{CR}}

\address[a]{Physics Department,
Florida State University,
Tallahassee, FL 32306-4350, U.S.A.}
\address[b]{Department of Astronomy and Theoretical Physics, 
Lund University,
SE-223 62 Lund, Sweden}

\fntext[SH]{sjh07@hep.fsu.edu}
\fntext[SQ]{squackenbush@hep.fsu.edu}
\fntext[LR]{reina@hep.fsu.edu}
\fntext[CR]{creuschle@hep.fsu.edu, christian.reuschle@thep.lu.se}

\begin{abstract}
  \nlox is a computer program for calculations in high-energy particle
  physics. It provides fully renormalized scattering matrix elements
  in the Standard Model of particle physics, up to one-loop accuracy
  for all possible coupling-power combinations in the strong and
  electroweak couplings, and for processes with up to six external
  particles.
\end{abstract}


\end{frontmatter}



{\bf PROGRAM SUMMARY}

\begin{small}
\noindent
{\em Program Title:} NLOX
\\
{\em Licensing provisions:} CC BY NC 3.0
\\
{\em Programming language:}
\cpp. \fortran interface available, and \fortran compiler required for
dependencies.
\\
%
{\em Required External Dependencies:}
\qcdloop 1.95, 
\oneloop 3.6 (available for download in utility tarball).
\\
{\em Required Compilers:} \gcc 4.6 or higher. Interface to certain
optional libraries requires \cpp11 support found in \gcc 4.7 or higher.
\\
{\em Operating System:}
Linux, MacOS.
\\




\end{small}

\setcounter{tocdepth}{1}

\section{Introduction}
\label{sec:intro}

The increasing level of precision of high-energy collider experiments
such as the Large Hadron Collider (LHC) has motivated the need for
theoretical predictions with accuracies at the percent level.  In
the high-energy regime of experiments like the LHC, cross sections
and branching ratios for elementary particle physics processes can be
derived from perturbative calculations within the Standard Model (SM)
and can be predicted with increasing theoretical accuracy the higher
the achievable perturbative order in theoretical calculations.

The last fifteen years have seen an impressive effort to move
perturbative calculations of scattering amplitudes for collider
physics to a new level of accuracy and to provide automatic tools for
their calculation. Theoretical predictions for processes of relevance to
precision physics at high-energy colliders have been pushed to include 
the next-to-next-to-leading order (NNLO) of strong (QCD) corrections and
the next-to-leading order (NLO) of electroweak (EW) corrections. If 
the effect of NLO QCD corrections on most collider processes is 
typically the dominant one, for several processes the effect of adding 
NLO EW corrections, hence considering mixed NLO QCD and EW 
corrections, can be as sizable as NNLO QCD corrections. Moreover, as
EW corrections become typically more prominent at high energies, at 
the energy regime explored by \eg Run II of the LHC the inclusion of 
NLO EW corrections on top of NLO and NNLO QCD corrections is
mandatory. 

Indeed, the problem of providing the NLO QCD and EW corrections for SM
$2\rightarrow 3$ and $2\rightarrow 4$ processes has been largely
tackled by dedicated calculations, and several automated frameworks
have been created that can push NLO SM calculations to even higher
final-state
multiplicities~\cite{Berger:2008sj,Alwall:2014hca,Cullen:2011ac,Cullen:2014yla,Cascioli:2011va,
  Actis:2016mpe,vanHameren:2009vq,Actis:2012qn,Badger:2012pg,Biedermann:2017yoi}.
A collection of NLO QCD cross sections for selected processes at
hadron colliders has been also made available in the context of the
\mcfm parton-level Monte Carlo program~\cite{mcfm8}, whose recent
version also includes NLO EW corrections for a small number of
processes~\cite{Campbell:2016dks}. Furthermore, methods to match
fixed-order NLO calculations with parton-shower Monte Carlo event
generators have been made available in several frameworks (\eg
\mgfive~\cite{Alwall:2014hca}, \powhegbox~\cite{Alioli:2010xd},
\sherpa~\cite{Gleisberg:2008ta},
\herwigseven~\cite{Bellm:2015jjp,Bellm:2017bvx}),
some of which follow standardized interface procedure between one-loop
calculations and Monte Carlo tools, as \eg discussed in
Refs.~\cite{Binoth:2010xt,Alioli:2013nda}. 

One-loop amplitudes are one important component of NLO QCD and EW
calculations, where they enter the virtual corrections to scattering
matrix elements, and the construction of efficient automatized
\textit{One-Loop Providers} (OLP) has played a major role in the field
of NLO QCD and EW studies in recent years. Traditional calculational
techniques of scattering amplitudes based on a Feynman diagram
approach can prove themselves quite inefficient for high-multiplicity
processes, unless special care is paid to their optimization.  With
the aim of reducing the complexity and improving the efficiency of
one-loop QCD and EW calculations, several OLP have been based on new
techniques, from unitarity-based
methods~\cite{Bern:1994zx,Bern:1994cg,Britto:2004nc,Ossola:2006us,
  Ellis:2007br,Giele:2008ve,Ellis:2008ir,vanHameren:2009dr}, to
improved diagrammatic techniques and recursion relations
~\cite{Cascioli:2011va,vanHameren:2009vq,Actis:2012qn,Weinzierl:2005dd,
  Becker:2012aqa,Reuschle:2013qna,Becker:2010ng}, as well as numerical
methods~\cite{Nagy:2003qn,Becker:2010ng,Nagy:2006xy,Gong:2008ww,
  Carter:2010hi,Becker:2012aqa,Becker:2012nk}. Traditional as well as
new methods have been implemented in several of the OLP that have been
largely used for LHC physics, \eg \blackhat~\cite{Berger:2008sj},
\vbfnlo~\cite{Arnold:2008rz,Arnold:2011wj,Baglio:2014uba},
\helaconeloop~\cite{vanHameren:2009dr,Bevilacqua:2011xh},
\madloop~\cite{Hirschi:2011pa} (embedded in
\mgfive~\cite{Alwall:2014hca}),
\gosam~\cite{Cullen:2011ac,Cullen:2014yla},
\openloops~\cite{Cascioli:2011va}, \njet~\cite{Badger:2012pg} , and
\recola~\cite{Actis:2016mpe,Actis:2012qn}.
Some of these codes provide pre-generated code for matrix elements of
parton-level processes, others allow to generate parton-level matrix
elements from scratch. They all can calculate one-loop QCD
corrections, while one-loop QCD and EW corrections have been fully
included in the public version of \mgfive, \openloops, and \recola.
Recent calculations of NLO QCD and EW corrections to important SM
processes obtained in these frameworks include the production of a
Higgs boson with a pair of on-shell~\cite{Frixione:2015zaa} or
off-shell top quarks~\cite{Denner:2016wet}, of $t\bar{t}$ plus
jets~\cite{Gutschow:2018tuk}, of EW vector bosons with
jets~\cite{Kallweit:2014xda,Kallweit:2015dum,Chiesa:2015mya} or
photons~\cite{Greiner:2017mft}, of diphotons plus
jets~\cite{Chiesa:2017gqx}, of a pair of off-shell EW
bosons~\cite{Kallweit:2017khh}, of $WWW$~\cite{Schonherr:2018jva}, and
of four on-shell top-quarks or of top pairs with a $W$
boson~\cite{Frederix:2017wme}.

\nlox, the program described in this article, is a new program for the
automated computation of one-loop QCD and EW corrections in the
Standard Model, which has recently been used in the computation of QCD
and EW corrections to $Z+b$-jet
production~\cite{Figueroa:2018chn,Figueroa:2018aqv}, 
and further partook in a technical comparison of tools for the
automation of NLO EW calculations~\cite{Bendavid:2018nar}.  A
non-public predecessor of \nlox has been available in the past, to
calculate one-loop QCD corrections to selected
processes~\cite{Reina:2011mb,Reina:2012lfa}.  \nlox has been
extensively expanded, and the current version of \nlox provides fully
renormalized QCD and EW one-loop corrections to SM processes at the squared-amplitude 
level, for all the possible QCD and EW mixed coupling-power
combinations to a certain parton-level process up to one-loop
accuracy, including the full mass dependence on initial- and
final-state particle masses.\footnote{ Currently only top and bottom
  quarks can be treated as massive, while the first four quark flavors as
  well as all leptons are treated as massless.}  Based on a
Feynman-diagram approach, with optimized parsing and storing of
recurrent building blocks,\nlox has been developed with the intent of
providing one-loop QCD and EW corrections in the SM, while maintaining
the flexibility to be further extended.

The generation and evaluation of one-loop QCD and EW corrections of a
large variety of $2\rightarrow 3$ and $2\rightarrow 4$ SM processes
has been thoroughly tested and, with this paper, we are releasing a
description of the code functionalities, archives of pre-generated
process code, and the core code to evaluate and interface to the
pre-generated process code.

The conventions used in building renormalized one-loop QCD and EW
amplitudes with \nlox are summarized in Sec.~\ref{sec:conventions},
while the \nlox workflow and code generation is briefly illustrated in
Sec.~\ref{sec:workflow}.  The \tred tensor-reduction library, an
essential part of the \nlox package, is described in
Sec.~\ref{sec:tred}.  Sec.~\ref{sec:using-nlox} provides some
practical instructions on how to download, install, and use the
current public version of \nlox. Finally, in
Sec.~\ref{sec:benchmarks} we report benchmark results for selected
processes. A brief summary is provided in Sec.~\ref{sec:summary}. In
the two appendices we collect more details about renormalization and
tensor reduction in \nlox.

\section{Conventions}
\label{sec:conventions}

\nlox computes the \lo (LO) and one-loop \nlo (NLO) contributions to a certain
subprocess at the level of the UV renormalized, color- and 
helicity-summed squared amplitude, in the Stueckelberg-Feynman gauge.

\subsection{Dimensional Regularization}
\label{sec:dimreg}

In \nlox ultraviolet (UV) and infrared (IR) singularities are
regularized using $d$-dimensional regularization (with
$d=4-2\epsilon$, $|\eps|\ll1$).
UV singularities are renormalized (see Sec.~\ref{sec:renorm}), while IR singularities are reported
in terms of the Laurent coefficients of the corresponding
$1/\epsilon^2$ and $1/\epsilon$ poles.
By default, and the only currently supported choice, \nlox uses a
variant of the t'Hooft-Veltman (HV) scheme~\cite{tHooft:1972tcz}. In
the HV scheme, Lorentz indices belonging to "external" states, that
is, those not belonging to a loop, are kept as 4-dimensional. Lorentz indices belonging to ``internal'' states are
promoted to be $d$-dimensional. The algebra of Dirac gamma matrices,
\ie $\{\gamma^\mu,\gamma^\nu\}=2g^{\mu \nu}$, is preserved
accordingly, with $g^\mu_\mu=d$ if $\mu$ belongs to an internal state.
However, this does not specify what to do with axial currents and
$\gamma^5$. The original choice of HV, preserving it as a
4-dimensional object, is unwieldy but yields the correct result for
triangle anomalies. Instead the choice in \nlox is to preserve the algebraic
property that $\gamma^5$ anticommutes with all gamma matrices,
$\{\gamma^5,\gamma^\mu\}=0$, which is consistent with the derivation
of the EW counterterms used by \nlox. To obtain the correct results
from anomalous triangle diagrams thus requires some care in the
ordering of gamma matrices in traces, which it is handled in our
scripts by a reading-point prescription (see \eg Ref.~\cite{Kreimer:1993bh}).

\subsection{Amplitudes and Mixed Expansions}
\label{sec:mixedexpansions}

\nlox organizes SM parton-level amplitudes as an expansion in the
strong ($g_s$) and electromagnetic ($g_e$) couplings.  Given a
process, which at \lo is defined by tree amplitudes with $n$ external
particles, \nlox calculates the tree and UV renormalized one-loop
contributions to the total unpolarized, color- and helicity-summed
squared amplitude, which up to these contributions is given by
\begin{align}
\label{eq:ampsquared}
\left|\amp_n\right|^2=
\left|\amp_n^{(0)}\right|^2
\,+\,\,
2\,\mathrm{Re}\left(\amp_n^{(0)*}\amp_n^{(1)}\right)\,,
\end{align}
where $\amp_n^{(0)}$ and $\amp_n^{(1)}$ denote the tree-level and
one-loop $n$-particle amplitudes, whose mixed coupling-power
expansions in terms of $g_s$ and $g_e$ read
\begin{align}
\label{eq:amp0and1}
\amp_{n}^{(0)}
\,\,&=
\hspace{-1ex}
\sum\limits_{\substack{i,j\,\geq\,0\\i+\!j=n-2}}
\hspace{-1ex}
g_s^i\,g_e^j\,\amp_{n}^{(0)(i,j)}
\,\,=
\hspace{-1.7ex}
\sum\limits_{\substack{0\,\leq\,i\,\leq\,n-2}}
\hspace{-1.7ex}
g_s^i\,g_e^{n-2-i}\amp_{n}^{(0)(i)}
\,,\\
\amp_{n}^{(1)}
\,\,&=
\hspace{0.25ex}
\sum\limits_{\substack{i,j\,\geq\,0\\i+\!j=n}}
\hspace{0.25ex}
g_s^i\,g_e^j\,\amp_{n}^{(1)(i,j)}
\,\,=
\hspace{-0.5ex}
\sum\limits_{\substack{0\,\leq\,i\,\leq\,n}}
\hspace{-0.5ex}
g_s^i\,g_e^{n-i}\amp_{n}^{(1)(i)}
\,.
\end{align}
Notice that, making use of the fact that for any SM tree and one-loop
amplitude we have $j+i=n-2$ and $j+i=n$ respectively, we have labeled
the order-by-order terms in the expansion,
$\amp_{n}^{(0)(i)}$ and $\amp_{n}^{(1)(i)}$, by the power of $g_s$
only.  It follows that the one-loop amplitudes that can be generated
from a tree amplitude $\amp_n^{(0)(i)}$ are either
$\amp_n^{(1)(i+2)}$, by inserting loop particles coupling through a
QCD interaction, or $\amp_n^{(1)(i)}$, by inserting loop particles
coupling through an EW interaction.  The mixed expansions of the \lo
and \nlo terms in Eq.~(\ref{eq:ampsquared}) can then be written as
\begin{align}
\label{eq:borncorrespondence}
\left|\amp_n^{(0)}\right|^2
\,\,&=\,
\hspace{-5ex}
\sum\limits_{\substack{0\,\leq\,i^\prime\!\!,i\,\leq\,n-2\\0\,\leq\,i^\prime\!\!+i\,\leq\,2(n-2)}}
\hspace{-5ex}
g_s^{i^\prime\!\!+i}
g_e^{2(n-2)-i^\prime\!\!-i}
\left(\amp_n^{(0)(i^\prime)}\right)^{\!*}\!\left(\amp_n^{(0)(i)}\right)\,,\\
\label{eq:virtcorrespondence}
2\,\mathrm{Re}\left(\amp_n^{(0)*}\amp_n^{(1)}\right)
\,\,&=\,
\hspace{-7ex}
\sum\limits_{\substack{0\,\leq\,i^\prime\leq\,n-2\,,\;0\,\leq\,i\,\leq\,n\\0\,\leq\,i^\prime\!\!+i\,\leq\,2(n-1)}}
\hspace{-7ex}
g_s^{i^\prime\!\!+i}
g_e^{2(n-1)-i^\prime\!\!-i}
\,2\,\mathrm{Re}\left(\left(\amp_n^{(0)(i^\prime)}\right)^{\!*}\!\left(\amp_n^{(1)(i)}\right)\right)\,.
\end{align}
The results of \nlox are reported as coefficients $a_0$ and $c_0$,
$c_1$, $c_2$ of Laurent series in $\eps$, such that\footnote{
  Notice that we define the Laurent coefficients to include all
  associated coupling powers of $g_s$ and $g_e$ and to include the
  common factor of $1/(2\pi)$ from the loop integration.}
\begin{align}
\label{eq:lolaurent}
g_s^{i^\prime\!\!+i}
g_e^{2(n-2)-i^\prime\!\!-i}
\left(\amp_n^{(0)(i^\prime)}\right)^{\!*}\!\left(\amp_n^{(0)(i)}\right)
&\,=
\,a_0^{(i^\prime\!,i)}\;+\;\order{\eps}\,,\\
\label{eq:nlolaurent}
g_s^{i^\prime\!\!+i}
g_e^{2(n-1)-i^\prime\!\!-i}
\,2\,\mathrm{Re}\left(\left(\amp_n^{(0)(i^\prime)}\right)^{\!*}\!\left(\amp_n^{(1)(i)}\right)\right)
&\,=
\,S_\eps\,
\bigg(\frac{c_2^{(i^\prime\!,i)}}{\eps^2}+\frac{c_1^{(i^\prime\!,i)}}{\eps}+c_0^{(i^\prime\!,i)}\bigg)\;+\;\order{\eps}
\,,
\end{align}
with $S_\eps=(4\pi)^\eps/\Gamma(1-\eps)$. The previous expansion can
be easily converted into an expansion in $\alpha_s=g_s^2/(4\pi)$ and
$\alpha_e=g_e^2/(4\pi)$, where the \lo contributions in
Eq.~(\ref{eq:borncorrespondence}) contribute to
$\alpha_s^x\alpha_e^{n-2-x}$, with $0\!\leq\!x\!\leq\!n\!-\!2$, while
the one-loop \nlo contributions in Eq.~(\ref{eq:virtcorrespondence})
contribute to $\alpha_s^x\alpha_e^{n-1-x}$, with
$0\!\leq\!x\!\leq\!n\!-\!1$, and where in both types of contributions
various amplitude-level $(i^\prime\!,i)$ configurations contribute to
the same $x=(i^\prime\!+i)/2$ at the squared-amplitude level.

It is instructive at this point to give a quick example, in order to
illustrate the subtleties of mixed coupling expansions.  Consider
two-parton production at the LHC, \ie $pp\rightarrow jj$, with
$p\ni\{\mathrm{quarks},g\}$ and $j\ni\{\mathrm{quarks},g\}$ (see the
benchmark processes in Sects.~\ref{sec:ppttbar} and~\ref{sec:ppjj}).
The possible subprocesses are all those with amplitudes with four
quarks, with two quarks and two gluons, and with four gluons.  The
mixed expansions on the amplitude and squared-amplitude level for this
process are depicted in Figs.~\ref{fig:cpflowampppjj}
and~\ref{fig:cpflowamp2ppjj} respectively:
\begin{figure}
\centering
\begin{subfigure}[t]{0.95\textwidth}
\begin{center}
\hspace{-1.5ex}
\includegraphics[scale=0.75]{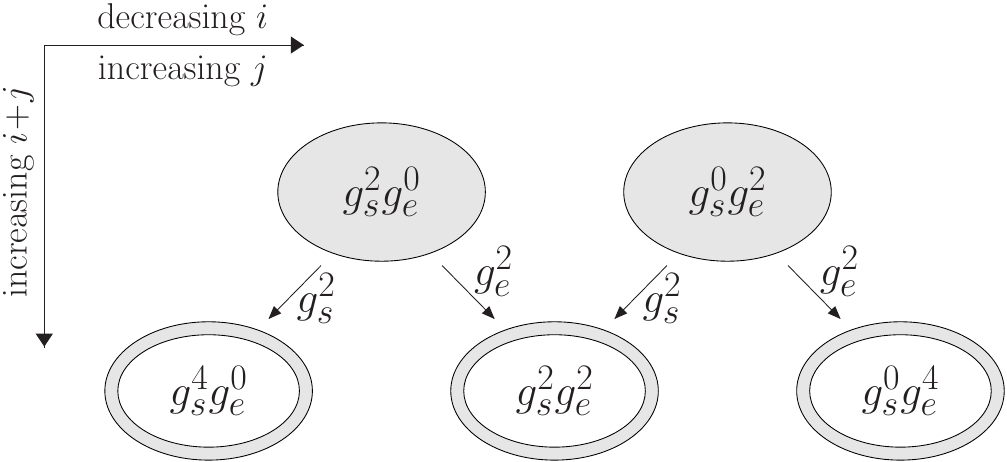}
\caption{\label{fig:cpflowampppjj} \small We consider the
  coupling-power combinations $g_s^ig_e^j$ at the amplitude level.
  From left to right we have increasing $i$ / decreasing $j$ in steps
  of 2. From top to bottom the total order $i\!+\!j$ increases in
  steps of 2. The upper row depicts all possible coupling-power
  combinations for the \lo contributions, with $i\!+\!j\!=\!n\!-\!2$,
  while the lower row depicts all possible coupling-power combinations
  for the higher-order corrections of one loop order higher, with
  $i\!+\!j\!=\!n$.}
\end{center}
\end{subfigure}

\vspace{1ex}
\begin{subfigure}[t]{0.95\textwidth}
\begin{center}
\hspace{-1.5ex}
\includegraphics[scale=0.75]{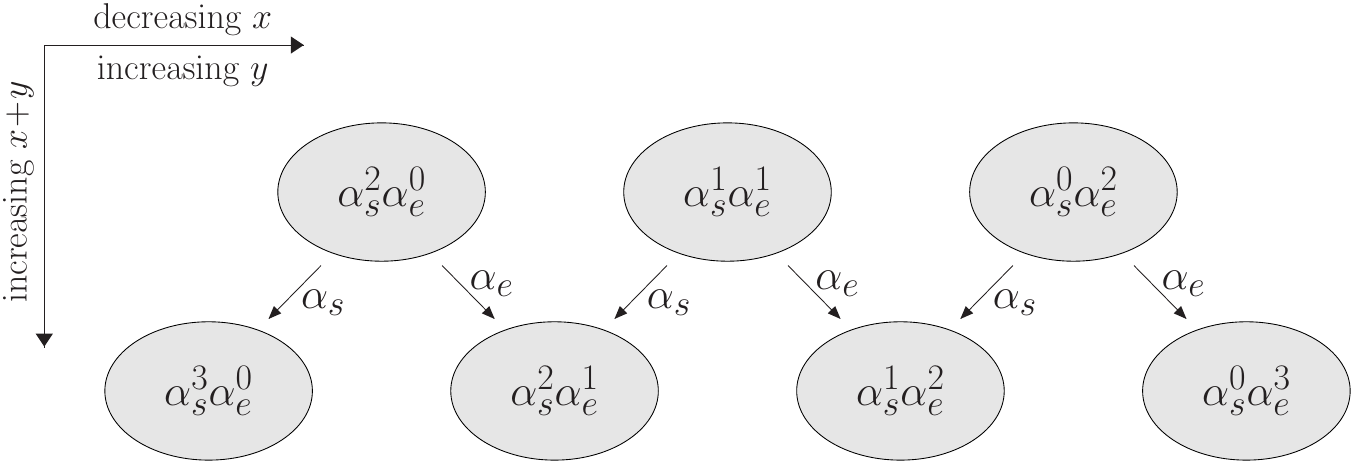}
\caption{\label{fig:cpflowamp2ppjj} \small {We consider} the
  coupling-power combinations $\alpha_s^{x}\alpha_e^{y}$ at the
  squared-amplitude level. From left to right we have increasing $x$ /
  decreasing $y$ in steps of 1. From top to bottom the total order
  $x\!+\!y$ increases in steps of 1.  The upper row depicts all
  possible coupling-power combinations for the \lo contributions, with
  $x\!+\!y\!=\!n\!-\!2$, while the lower row depicts all possible
  coupling-power combinations for the higher-order corrections of one
  order higher, with $x\!+\!y\!=\!n\!-\!1$.}
\end{center}
\end{subfigure}
\caption{\label{fig:cpflow} \small Coupling-power flow chart for
  two-parton production at the LHC, for which the \lo contributions
  are defined by tree amplitudes with $n=4$ external particles, at the
  level of (a) the amplitude and (b) the squared amplitude. See the
  text for more details.}
\end{figure}
\begin{itemize}
\item At the amplitude level (see Fig.~\ref{fig:cpflowampppjj}), we
  note that the $g_s^4g_e^0$ one-loop contribution is obtained solely
  from $g_s^2$ corrections to the $g_s^2g_e^0$ \lo contribution, and
  that the $g_s^0g_e^4$ one-loop contribution is obtained solely from
  $g_e^2$ corrections to the $g_s^0g_e^2$ \lo contribution. However,
  the $g_s^2g_e^2$ one-loop contribution cannot be classified solely
  as a QCD or EW one-loop correction to a particular \lo contribution,
  as it contains $g_s^2$ corrections to $g_s^0g_e^2$ and $g_e^2$
  corrections to $g_s^2g_e^0$ at the same time, where some of those
  corrections are being represented by the same diagrams, which do not
  need to be counted twice.  Looking specifically at the four-quark
  subprocesses of the type $q\bar{q}\rightarrow q'\bar{q}'$, the two
  \lo contributions are either through gluon exchange or $Z$/$\gamma$
  exchange, and for example the one-loop box correction which consists
  of, say, $qgq'Z$ is only generated once, although being at the same
  time a $g_s^2$ correction to the \lo $Z$/$\gamma$-exchange
  contribution as well as a $g_e^2$ correction to the \lo
  gluon-exchange contribution.
\item At the squared-amplitude level (see
  Fig.~\ref{fig:cpflowamp2ppjj}), the possible \lo contributions are
  $\alpha_s^2\alpha_e^0$ for all of the subprocesses, as well as
  $\alpha_s^1\alpha_e^1$ and $\alpha_s^0\alpha_e^2$ for only the
  four-quark subprocesses. We note that the $\alpha_s^2\alpha_e^0$ \lo
  contributions result solely from squaring the corresponding
  amplitude-level $g_s^2g_e^0$ \lo contributions, the order
  $\alpha_s^0\alpha_e^2$ \lo contributions results solely from
  squaring the corresponding amplitude-level $g_s^0g_e^2$ \lo
  contributions, while the $\alpha_s^1\alpha_e^1$ \lo contributions
  are the result of interfering the amplitude-level contributions of
  different amplitude-level coupling-power combinations, \ie
  interfering the $g_s^2g_e^0$ tree diagrams with the $g_s^0g_e^2$
  tree diagrams consisting of the same external configurations.  The
  one-loop \nlo contributions to the contributions in the lower row of
  Fig.~\ref{fig:cpflowamp2ppjj} are obtained by interfering the
  contributions from the upper row of Fig.~\ref{fig:cpflowampppjj}
  with the contributions from the lower row of
  Fig.~\ref{fig:cpflowampppjj}, as formulated generically in
  Eq.~(\ref{eq:virtcorrespondence}).  Further note, although the
  $\alpha_s^1\alpha_e^1$ \lo contribution, from interfering the \lo
  gluon-exchange and $Z$/$\gamma$-exchange diagrams, is identically
  zero, due to color arguments, the subset of the
  $\alpha_s^2\alpha_e^1$ one-loop \nlo contribution that can be
  reached from the $\alpha_s^1\alpha_e^1$ \lo contribution by
  advancing one power in $\alpha_s$ is not zero.
\end{itemize}

\subsection{Renormalization}
\label{sec:renorm}

UV renormalization in \nlox is carried out by means of counterterm
(CT) diagrams.  Upon generation of diagrams, also all possible CT
diagrams are generated, and sorted by coupling powers.  The CT
diagrams with the same coupling power as a certain set of one-loop
diagrams get eventually associated with that set.

The renormalization constants in terms of which the QCD UV
counterterms are formulated are derived in a mixed renormalization
scheme: a modified on-shell scheme is used for the wave-function and
mass renormalization of massive quarks, while the $\msbar$ scheme is
used for massless quarks and gluons, where, however, in the latter
case heavy-quark-loop contributions are decoupled by subtracting them
at zero momentum~\cite{Collins:1978wz,Nason:1987xz}.\footnote{We provide an extension of the schemes presented in
  Ref.~\cite{Collins:1978wz,Nason:1987xz}, applicable \eg also in the
  case of a massive five-flavor scheme with $\msbar$ PDF (see
  Ref.~\cite{Figueroa:2018chn}). More details are provided
  in~\ref{app:renormalization}.}

The renormalization constants in terms of which the EW UV counterterms
are formulated are derived in the on-shell renormalization scheme as
described in Ref.~\cite{Denner:1991kt}\,%
\footnote{ At difference from Ref.~\cite{Denner:1991kt} we use
  dimensional regularization both for UV and IR divergences, including
  soft singularities which in Ref.~\cite{Denner:1991kt} are regulated
  by a photon mass.}, or, in the presence of potential resonance
channels, in the complex-mass scheme~\cite{Denner:2005fg}, where the
choice in \nlox is to expand self-energies with complex squared
momenta around real squared momenta.  As EW input scheme choices \nlox
provides both the $\alpha(0)$ and the $G_\mu$ EW input
schemes~\cite{Hollik:1988ii,Denner:1991kt, Dittmaier:2001ay,
  Andersen:2014efa}. Per default the $\alpha(0)$ EW input scheme is
used. 

More details are provided in~\ref{app:renormalization}.

\section{Overview on Workflow and Code Generation}
\label{sec:workflow}

\nlox utilizes \qgraf~\cite{Nogueira:1991ex},
\form~\cite{Vermaseren:2000nd,Kuipers:2012rf}, and \python to
algebraically generate \cpp code for the virtual QCD and EW one-loop
contributions to a certain process in terms of one-loop
tensor-integral coefficients.  The tensor-integral coefficients are
calculated recursively at runtime through standard reduction methods
by the \cpp library \tred, an integral part of \nlox

In this chapter, we give a brief description of how \nlox generates 
the code and data for a process. As the current release of \nlox 
contains pre-generated process code and the \tred library, we defer a 
more complete discussion for a future release of \nlox, which will 
enable the user to generate process code on their own.

\subsection{Algebraic Processing}
\label{sec:algebra}

\nlox begins with text-based model files containing the fields and
vertices of the model under consideration. The current version
supports the Standard Model with QCD and EW corrections as described
in Sec.~\ref{sec:conventions} and \ref{app:renormalization}.  Such
\nlox model is then parsed into a model file for
\qgraf~\cite{Nogueira:1991ex}, which produces a list of diagrams for
further processing.  \python scripts take the \qgraf output and
perform substitutions and simplifications to prepare for further
processing by \form scripts. In this step diagrams of the wrong
coupling order, that is those not corresponding to the requested \lo
or one-loop \nlo contributions are discarded.  Finally
\form~\cite{Vermaseren:2000nd,Kuipers:2012rf} is used to perform
further algebraic manipulations and substitutions.

Three main ingredients into the algebraic processing are the treatment
of color, the tensor decomposition of one-loop tensor-integrals and 
the utilization of Standard Matrix Elements, which we will briefly
discuss below.

\paragraph{Color Treatment}
At the amplitude level, the \form scripts of \nlox identify the
various color structures, \ie products of color matrices, simplify
them, and cache them whilst identifying identical ones, such that only
a limited set of unique color strings is kept.  At the
squared-amplitude level, the various unique color structures are
interfered and numerically evaluated, as the interference terms of the
various diagrams that contribute to the requested coupling order are
evaluated.

\paragraph{Tensor Decomposition}

Loop diagrams will contain integrals over the loop momentum. The \form
scripts of \nlox detect these loops, and use a standard decomposition
of a tensor integral onto a basis of Lorentz
structures~\cite{Denner:1991kt,Denner:2005nn}.  What remains after
tensor decomposition are the one-loop tensor-integral coefficients and
external momenta of the process. The external momenta are folded into
the rest of the diagram, with the tensor coefficients remaining
symbolically during script processing, ultimately to be computed at
runtime by the tensor-reduction library \tred (see
Sec.~\ref{sec:tred}).

\paragraph{Standard Matrix Elements}

From a computing standpoint, the most challenging aspect of a
diagram-based approach is processing the many gamma matrix expressions
that appear in the calculation. To minimize this, strings of gamma
matrices (multiplied by momenta and external polarization vectors) in
a diagram are brought into a canonical order by anticommutation.
Later scripts identify them and reduce the code to unique structures,
hence improving efficiency. By an abuse of the standard terminology,
we call these unique strings of gamma matrices multiplied by momenta
and polarization vectors Standard Matrix Elements (SMEs).  After
diagram processing, the unique SMEs from a loop or tree amplitude are
interfered with the unique (conjugated) SMEs of a tree
amplitude. After simplification, a interfered pairs of SMEs, which we
will refer to as \textit{ISME} from here on, can only have
uncontracted Lorentz indices within a diagram if the indices belong to
different fermion lines.  We can safely take all remaining
$d$-dimensional indices to be 4-dimensional for SMEs in the HV
scheme.\footnote{ In the HV scheme (used by \nlox), these indices may
  be $d$- or 4-dimensional. They are only $d$-dimensional in the case
  where both fermion lines form part of a loop, which is necessarily
  4-point or higher and can contain only IR divergences owing to its
  limited rank.  In Ref.~\cite{Bredenstein:2008zb}, all rational terms
  of IR origin were proven to cancel, and therefore, for simplicity,
  we can take all remaining $d$-dimensional indices to be
  4-dimensional for SMEs, leading to a more efficient computation.  }.
Finally the ISMEs are computed by performing polarization/spin sums
and taking traces of the combined strings of gamma matrices (all in 4
dimensions\footnotemark[\value{footnote}])

At this stage we have color- and helicity-summed expressions stored
for each tree or loop diagram interfered with the sum of conjugated
tree diagrams, which we call an \texttt{xdiagram}, and which contains
references to the ISMEs and to the tensor coefficients in case of loop
diagrams, and expressions stored for the ISMEs themselves.

\subsection{\cpp Code and Process Data Generation}
\label{sec:codegeneration}

The final stage of process code generation consists of reading the
stored analytical \texttt{xdiagram} expressions, and turning them into
\cpp code to be compiled as well as raw data to be read at runtime. To
that end \nlox has a sophisticated \python script to parse \form
output. The script reads \texttt{xdiagram} expressions and ISMEs, and
processes the stored expressions.  During parsing, any quantity
recognized as a variable is identified to be used later in a list of
needed variables, to be passed by an interface code. These variables
can include constants, couplings, masses, momentum invariants,
denominators, and momentum-contracted epsilon tensors.  In the case of
the latter three, the script generates code to calculate them from
momenta and masses. In addition, the list of needed tensor
coefficients is compiled.

In one-loop diagrams we must keep terms of $O(\epsilon^0)$ through
$O(\epsilon^2)$ in case they multiply poles in
$\epsilon^{-1}$ or
$\epsilon^{-2}$ from the tensor coefficients, resulting in finite pieces.
A convenient solution from a \cpp coding standpoint is to encapsulate
the expression in a class called \polythree (see \ref{sec:poly3}),
which stores the expression as a three-term polynomial in $\epsilon$.

Turning the \form output into \cpp code directly can result in large
code sizes, a disadvantage of using a diagram-based approach,
especially in the case of \form where its strategy is to flatten
expressions to many small terms to be processed serially. However, the
advantage of the approach is that the expressions produced are very
regular. As such, we have implemented options for turning these
regular expressions into data stored in text files, to be read at run
time. Their computation then requires only simple loops over many
terms, both for the diagrams and ISMEs. Many identical expressions are
also identified at this stage. In our current release we only support
producing processes that have been turned to data to the full extent
possible.  Not only does this reduce code size, but it results in much
faster code.

The final result is a small \cpp code that computes with the following 
steps:
\begin{itemize}
\item Pass to the process the variables it needs, namely constants,
  couplings, masses, and momenta for a given phase-space point (PSP).
\item Calculate the ISMEs, whose results are stored in an array to be
  retrieved by the \texttt{xdiagram} calculating code.
\item Calculate the needed tensor coefficients, which we call
  \texttt{TI}s, stored in a \tred object (see Sec.~\ref{sec:tred}).
\item For each \texttt{xdiagram}, calculate the result, using the ISME
  and tensor coefficient results as input. Schematically these
  calculations are loops summing \texttt{prefactor*ISME*TI}, where
  each prefactor contains all the dependence of the diagram that is
  not an ISME or tensor coefficient, usually couplings, masses, and
  momentum invariants.
\end{itemize}

\section{Tensor Reduction and the \tred library}
\label{sec:tred}

The tensor-integral coefficients (see Sec.~\ref{sec:codegeneration}),
in terms of which the \cpp code for the virtual contributions to a
particular process is expressed, are calculated recursively at runtime
by the \cpp library \tred.  Several reduction techniques are available
to \tred, many of which are found in
Refs.~\cite{Passarino:1978jh,Denner:2005nn,Diakonidis:2008ij}.  From
here on we use \tred interchangeably to refer either to the
tensor-reduction library \tred or an object of the \tred class
(see \ref{sec:tred-class}).

\tred accumulates and stores a list of all tensor coefficients which
appear during the recursion and are needed by a particular process,
along with their dependencies.  Only needed coefficients are computed,
and no coefficient is computed more than once, making the reduction
process particularly efficient. The coefficient values are returned as
\polythree (see \ref{sec:poly3}).

\subsection{Functionality of \tred}
\label{sec:tredfunctionality}

As mentioned in Sec.~\ref{sec:algebra}, tensor integrals appearing in
Feynman diagrams are identified and tensor decomposed during diagram
processing by \form scripts. The tensor coefficients appearing in this
expansion are identified in \python scripts during source code
generation, and a list of needed coefficients is accumulated. This
list is loaded at runtime, and one-by-one a tensor coefficient is
requested of \tred. Each created tensor coefficient is stored in
\tred, and a pointer returned to the calling process for later
retrieval of its value.

Many schemes to reduce tensor one-loop integrals to combinations of
known scalar one-loop integrals have been described in the literature.
One approach to implement these schemes would be to compute the tensor
coefficients analytically and code the result into a library. Not even
accounting for the different kinematic limits of each tensor
coefficient, this would require hundreds of coded functions just to
cover 5-point integrals.  However, it may be more efficient and
general to do this \textit{numerically} at run-time by coding the
functions that determine one coefficient from others, which is the
approach that \tred take, thereby implementing several traditional
reduction schemes, many of which are found in
Refs.~\cite{Passarino:1978jh,Denner:2005nn,Diakonidis:2008ij}.

The dependencies of a given coefficient are different depending on the
chosen reduction method. Multiple methods, some overlapping, are
implemented in \tred. When a coefficient is requested during process
initialization, the following steps are taken:
\begin{itemize}
\item Check whether the coefficient is already stored. If so, just
  return a pointer to it. If not, build a new coefficient.
\item If a new coefficient is needed, for each active reduction
  method, determine the new coefficient's dependencies. If a
  dependency already exists, create and store a pointer to it for use
  by the new coefficient. If not, create \textit{that} coefficient,
  and so on, recursively.
\end{itemize}
Eventually for a given method and requested coefficient, this process
will terminate with coefficients that can be computed directly,
usually a scalar coefficient. This process guarantees that all needed
coefficients are available, but no unneeded coefficients are ever
created or computed.  The coefficient objects (and pointers to
dependencies) are created once at runtime. From there, \tred is given
a PSP and asked to evaluate its stored coefficients. In \nlox a
process then is computed at that PSP using these coefficient values as
input.

The scalar integrals that stand at the end of the recursions are not
computed by the \tred library, as many libraries are already
available. Rather, interfaces to several external scalar libraries are
available in \tred: \qcdloop\,1~\cite{Ellis:2007qk} and
2~\cite{Carrazza:2016gav}, \looptools~\cite{Hahn:1998yk}, and
\oneloop~\cite{vanHameren:2010cp}.  In its current version the \nlox
package, provided through the \nloxutil collection of necessary
dependencies, contains \qcdloop\,1.95 and \oneloop 3.6 by default.

\subsection{Features of \tred}
\label{subsec:tredfeat}

\paragraph{Multiple Reduction Methods}

As of this note, the following reduction methods are implemented and 
used in \tred:
\begin{itemize}
\item 1- and 2-point integrals ($A$ and $B$ coefficients): Rather than
  use a standard reduction, these are computed directly, without
  numerical recursion, using the analytical formulas of
  Ref.~\cite{Denner:2005nn}. However, in the current implementation of
  the Standard Model used by \nlox and reduction methods available for
  higher-point integrals, the only $A$ coefficients appearing in the
  reduction are scalars and the higher-rank $A$ coefficients are not
  used.
\item 3- and 4-point integrals ($C$ and $D$ coefficients): Two
  reduction methods are implemented, Passarino-Veltman
  (PV)~\cite{Passarino:1978jh} and an alternate, similar reduction of
  Denner and Dittmaier (DD)~\cite{Denner:2005nn}, that uses modified
  Cayley matrices instead of Gram matrices.\footnote{Note that this
    scheme is denoted as PV$^\prime$ in Ref.~\cite{Denner:2005nn}.} As
  the Cayley matrices are singular when there are IR singularities,
  \tred requires that PV be available, and when such a singularity is
  detected \tred will disable the DD node and revert to PV. PV alone
  is enabled by default.
\item 5- and 6-point integrals ($E$ and $F$ coefficients): These types
  of nodes are different in that their reductions make use of the fact
  that only four independent momenta are needed to span a
  4-dimensional spacetime. Two reduction methods are implemented, one
  by Diakonidis et.~al.~\cite{Diakonidis:2008ij}, and one by Denner
  and Dittmaier~\cite{Denner:2005nn}. As the former method is limited
  to rank 3 integrals, we use the latter by default.
\end{itemize}
The \tred source code is extensible to other methods without much
modification beyond the actual implementation of the method, and we
anticipate adding others  in the future.

\paragraph{Stability Checks and Higher Precision}
\label{sec:stability}

$\epsilon$-pole parts of tensor coefficients tend to be much simpler
analytically than their corresponding finite parts. This is especially
true for IR-finite coefficients. A library of coefficients with
UV-only divergences exists inside \tred for the purposes of comparing
to values determined by the numerical reduction method -- this code is
simple and fast, so it can be used to check the numerical reduction
without significant effect on the runtime.  If a given coefficient
fails to match within a certain threshold, it \textit{and all its
  dependencies} are recomputed at a higher floating-point precision
level automatically. Three precision levels are attempted: double
precision (64-bit), long double precision (80-bit on typical modern
machines) and finally 128-bit (implemented by the \texttt{quadmath}
library). In addition, it is possible to request all coefficients at a
higher precision by requesting evaluation from \tred again before
feeding a new PSP, which may be useful if further checks are done at
the level of an interfaced process, where for instance the poles of a
renormalized virtual and real process are checked for
cancellation. If, after recomputation at the highest set
floating-point precision, a given coefficient still fails to match
within a certain precision, the one-loop contribution in question at
the PSP in question is deemed inaccurate, and \tred returns a
corresponding flag.

\paragraph{Kinematics Cache}

Invariants and various matrices, determinants, etc. are used (and
reused) during the numerical reduction. Rather than constantly
recompute them, for each PSP they are computed on first request and
stored for future use. The cache is cleared when the PSP is updated.

\section{Using \nlox}
\label{sec:using-nlox}

To access the host \texttt{URL} for the \nlox package,
please go to \url{http://www.hep.fsu.edu/~nlox}. For further
details on downloading and using \nlox, beyond what is described in 
the section at hand, please follow the instructions on the website.

\subsection{Components of the \nlox package}
\label{sec:nlox-components}

The current release of the \nlox package contains the following.
\begin{itemize}
\item \nlox: This first public release of \nlox consists of the \tred
  library, as well as the functionality to interface process archives,
  containing already generated process code.\footnote{ In a future
    release of \nlox, we will enable the user to generate process code
    on their own.}
\item \nloxutil: Scalar integrals are not computed by the \tred
  library, as many libraries are already available.  Rather,
  interfaces to external scalar libraries are available.
  \qcdloop\,1~\cite{Ellis:2007qk} and
  \oneloop~\cite{vanHameren:2010cp} are the current default and are in
  this release of the \nlox package provided through the \nloxutil
  collection of necessary dependencies, containing \qcdloop\,1.95 and
  \oneloop 3.6.\footnote{ Interfaces to
    \qcdloop\,2~\cite{Carrazza:2016gav} and
    \looptools~\cite{Hahn:1998yk} are also available, but have not
    been thoroughly tested and are not delivered with \nloxutil in the
    current release. In a future release of \nlox, including the
    capabilities to generate process code, \nloxutil will further
    contain compatible versions of \qgraf and \form.}
\item Process archives: Currently, already generated process code is
  provided through process archives, \ie tarballs containing
  pre-generated process code.\footnote{ Unless stated otherwise, all processes
    generated by \nlox allow for top and bottom quarks to be massive,
    while the first four quark flavors as well as all leptons are
    treated as massless.}
\end{itemize}

\begin{sloppypar}
A simple set of instructions on how to download and install the 
various components can be found on
\url{http://www.hep.fsu.edu/~nlox}.
\end{sloppypar}

\subsection{Interfacing Processes}
\label{sec:processinterface}

\nlox computes the \lo and one-loop \nlo contributions to a certain
subprocess at the level of the UV renormalized, color- and
helicity-summed squared amplitude and returns the values for the
Laurent coefficients defined in Eqs.~(\ref{eq:lolaurent})
and~(\ref{eq:nlolaurent}), which we briefly repeat
\footnote{ Note again, we define the Laurent coefficients to include
  all associated coupling powers of $g_s$ and $g_e$ and to include the
  common factor of $1/(2\pi)$ from the loop integration.}:
\begin{align}
\label{eq:lolaurent2}
g_s^{i^\prime\!\!+i}
g_e^{2(n-2)-i^\prime\!\!-i}
\left(\amp_n^{(0)(i^\prime)}\right)^{\!*}\!\left(\amp_n^{(0)(i)}\right)
&\,=
\;a_0^{(i^\prime\!,i)}\;+\;\order{\eps}\,,\\
\label{eq:nlolaurent2}
g_s^{i^\prime\!\!+i}
g_e^{2(n-1)-i^\prime\!\!-i}
\,2\,\mathrm{Re}\left(\left(\amp_n^{(0)(i^\prime)}\right)^{\!*}\!\left(\amp_n^{(1)(i)}\right)\right)
&\,=
\;S_\eps\,
\bigg(\frac{c_2^{(i^\prime\!,i)}}{\eps^2}+\frac{c_1^{(i^\prime\!,i)}}{\eps}+c_0^{(i^\prime\!,i)}\bigg)\;+\;\order{\eps}\,,
\end{align}
with $S_\eps=(4\pi)^\eps/\Gamma(1-\eps)$.
Given a process, of which the \lo contributions are defined by tree
amplitudes with $n$ external particles, in terms of an expansion in
$\alpha_s^x\alpha_e^y$, for a particular \lo Laurent coefficient
$a_0^{(i'\!,i)}$ with a combined power $(i'\!\!+\!i)=2x$ of $g_s$ in
the tree${}^*\!$/tree interference we read off the combined power of
$g_e$ to be $2y=(j'\!\!+\!j)=2(n\!-\!2)\!-\!(i'\!\!+\!i)$, whereas for
a particular one-loop \nlo Laurent coefficient $c_\eps^{(i'\!,i)}$
with a combined power $(i'\!\!+\!i)=2x$ of $g_s$ in the
tree${}^*\!$/one-loop interference we read off the combined power of
$g_e$ to be $2y=(j'\!\!+\!j)=2(n\!-\!1)\!-\!(i'\!\!+\!i)$.

Once \nlox and \nloxutil are installed, and the libraries in the
process archive for the process under consideration are compiled (see
previous section),
\begin{itemize}
\item the only file that needs to be included in the users main
  program is \texttt{nlox\_olp.h} for a \cpp program and
  \texttt{nlox\_fortran\_interface.f90} for a \fortran program,
\item the value for a particular \lo or \nlo Laurent coefficient of a
  particular subprocess can be retrieved by the function
  \texttt{NLOX\_OLP\_EvalSubProcess()}. This function is based on the
  standard BLHA~\cite{Alioli:2013nda}, with additional arguments to
  select the desired power of couplings.
\end{itemize}

\bigskip
In a \cpp program we have
\begin{itemize}
\item[] \texttt{NLOX\_OLP\_EvalSubProcess(\&isub, typ, cp, pp, \&next,
    \&mu, rval2, \&acc)}
\end{itemize}

with
\begin{itemize}
\item[] \texttt{isub}: Integer number specifying the ID of the
  subprocess in question.
\item[] \texttt{typ}: Character string specifying the interference
  type, \ie either \texttt{"tree\_tree"} or \texttt{"tree\_loop"}
  (note the double quotes).
\item[] \texttt{cp}: Character string specifying the coupling power of
  the interference type in question, in the format
  \texttt{"g$i'$e$j'$\_g$i$e$j$"}.
For example, for  
$u\bar{u}\rightarrow t\bar{t}$ 
the interference of the tree amplitude with the highest possible $g_s$ 
power of 2 (and a complementing $g_e$ power of 0) with the one-loop 
amplitude with the highest possible $g_s$ power of 4 (and a 
complementing $g_e$ power of 0) has \texttt{cp}=\texttt{"g2e0\_g4e0"}. 
\vspace{1ex}
\begin{itemize}
\item For a particular \texttt{isub} and a particular \texttt{typ},
  there are potentially multiple values of \texttt{cp} that contribute
  to the same coupling power at the squared-amplitude level, \ie in
  terms of an expansion in
  $\alpha_s^x\alpha_e^y= \alpha_s^{(i'\!+i)/2}\alpha_e^{(j'\!+j)/2}$.
  For \texttt{typ}=\texttt{"tree\_loop"} they need to be retrieved
  separately and summed, \eg \texttt{cp}=\texttt{"g2e0\_g2e2"} and
  \texttt{cp}=\texttt{"g0e2\_g4e0"} for
  $u\bar{u}\rightarrow t\bar{t}$.  For
  \texttt{typ}=\texttt{"tree\_tree"} each yields the same result, \eg
  \texttt{cp}=\texttt{"g2e0\_g0e2"} or
  \texttt{cp}=\texttt{"g0e2\_g2e0"} for
  $u\bar{u}\rightarrow t\bar{t}$.
\item All available values for \texttt{isub}, \texttt{typ} and
  \texttt{cp}, for a particular process in question, are listed in the
  \texttt{SUBPROCESSES} file in the corresponding process archive.
\end{itemize}
\item[] \texttt{mu}: Double precision number specifying the scale, in
  units of GeV.
\item[] \texttt{next}: Integer determining the number of external
  particles of the subprocess in question.
\item[] \texttt{pp}: Array of double precision numbers, with dimension
  $5\cdot \texttt{next}$, which specifies the PSP
  $\{p_1,p_2\}\rightarrow\{p_3,...,p_{\texttt{next}}\}$, in units of
  GeV, and is of the form \texttt{pp} = \texttt{[ $p_{1,t}$\,,
    $p_{1,x}$\,, $p_{1,y}$\,, $p_{1,z}$\,, $m_1$\,, $p_{2,t}$\,,
    $p_{2,x}$\,, ...\,]}.
\item[] \texttt{rval2}: Returned array of double precision numbers,
  with dimension 3, which contains the results for the Laurent
  coefficients of order $(i'\!,i)$ in Eqs.~(\ref{eq:lolaurent2})
  and~(\ref{eq:nlolaurent2}), evaluated at \texttt{pp}, and
  \vspace{1ex}
\begin{itemize}
\item for \texttt{typ} = \texttt{"tree\_tree"} is of the form
  \texttt{rval2[0,1,2]} =
  \texttt{[$\,0\,$,$\,0\,$,$\,a_0^{(i^*\!,i)}\,$]},
\item for \texttt{typ} = \texttt{"tree\_loop"} is of the form
  \texttt{rval2[0,1,2]} = \texttt{[$\,c_2^{(i^*\!,i)}$,
    $c_1^{(i^*\!,i)}$, $c_0^{(i^*\!,i)}\,$]}.
\end{itemize}
\vspace{1ex}
\item[] \texttt{acc}: Always reports the integer 0 in this release.
\end{itemize}

\bigskip
In a \fortran program we have
\begin{itemize}
\item[] \texttt{NLOX\_OLP\_EvalSubProcess(isub, typ, ltyp, cp, lcp,
    pp, next, mu, rval2, acc)}
\end{itemize}

where 
\begin{itemize}
\item[] in addition to the above there are two additional arguments,
  \texttt{ltyp} and \texttt{lcp}, \ie integer numbers specifying the
  number of characters in the character strings \texttt{typ} (always
  \texttt{ltyp}=\texttt{9}) and \texttt{cp} (most likely always
  \texttt{lcp}=\texttt{9}) respectively. Where arrays in \cpp start
  with element \texttt{0}, arrays in \fortran start with element
  \texttt{1}:
  \texttt{pp[0,1,...,5$\cdot$next-1]}$\rightarrow$\texttt{pp(1,2,...,5$\cdot$next)}
  and \texttt{rval2[0,1,2]}$\rightarrow$\texttt{rval2(1,2,3)}.
\end{itemize}

\bigskip
\begin{sloppypar}
  A simple set of instructions on how to interface to the process
  archives, which further point to the example programs that are
  currently provided with each process archive, can be found on
  \url{http://www.hep.fsu.edu/~nlox}.
\end{sloppypar}

\section{Benchmarks}
\label{sec:benchmarks}

In this section we compare PSP
results 
of \nlox for selected processes against another program, 
\recola~\cite{Actis:2016mpe}. Comparisons to other codes can be found
in~\cite{honeywell:2017phd}.
%
In addition to serving as a useful check, this section can serve as a
useful reference for those attempting to test the installation and use
of a process archive. We also give \nlox time benchmarks for each of
the processes in this section.

\subsection{General Setup}

The results presented in this section are obtained calculating EW
corrections using the complex mass scheme and defining renormalized EW
parameters in the $\alpha(0)$ input scheme (see
Sect.~\ref{sec:renorm} and~\ref{sec:ewrenorm}). All particles have zero
width unless otherwise specified. We use a diagonal CKM
matrix. All leptons and the four lightest quarks are considered
massless, and all other parameters are listed in
Table~\ref{tab:params}. Their values are arbitrarily chosen just for
the sake of benchmarking, and correspond to values used \eg
already in~\cite{honeywell:2017phd} to compare with other packages such as \gosam{}
(see Ref.~\cite{honeywell:2017phd} for more details).
The same setup is also used as \nlox default setup, and will have to
be adapted to specific calculations as needed by modifying the choice
of input parameters as further explained on 
\url{http://www.hep.fsu.edu/~nlox}.

\begin{table}[H]
\centering
\begin{tabular}{l r|l r}
$\alpha$ & 1 & $\alpha_s$ & 1 \\
$m_b$ & 4.2~GeV & $m_t$ & 171.2~GeV \\
$m_W$ & 80.376~GeV & $m_Z$ & 91.1876~GeV \\
$m_H$ & 125~GeV & $\mu$ & 1000~GeV
\end{tabular}
\caption{Input values used in this chapter's comparisons, and the 
defaults for \nlox.}
\label{tab:params}
\end{table}
For simplicity we test all processes with a common set of PSPs,
obtained for a center-of-mass energy $\sqrt{s}=1$~TeV.  For a given
multiplicity, these PSPs only depend on the masses of the external
particles. We list them in Tables~\ref{tab:psp00tt}
through~\ref{tab:psp000000} of~\ref{sec:psp}.  According to
Table~\ref{tab:params}, the renormalization
scale is also chosen to be $\mu=1$~TeV.

\subsection{Processes}
\label{subsec:processes}

For all processes, we report PSP results for all or part of the
individual channels contributing to the process in terms of the
Laurent coefficients of the expansions in
Eqs.~(\ref{eq:lolaurent})/(\ref{eq:lolaurent2})
and~(\ref{eq:nlolaurent})/(\ref{eq:nlolaurent2}).  As explained above,
given a process, of which the \lo contributions are defined by tree
amplitudes with $n$ external particles, in terms of an expansion in
$\alpha_s^x\alpha_e^y$, for a particular \lo Laurent coefficient
$a_0^{(i'\!,i)}$ with a combined power $(i'\!\!+\!i)=2x$ of $g_s$ in
the tree${}^*\!$/tree interference we read off the combined power of
$g_e$ to be $2y=(j'\!\!+\!j)=2(n\!-\!2)\!-\!(i'\!\!+\!i)$, whereas for
a particular one-loop \nlo Laurent coefficient $c_\eps^{(i'\!,i)}$
with a combined power $(i'\!\!+\!i)=2x$ of $g_s$ in the
tree${}^*\!$/one-loop interference we read off the combined power of
$g_e$ to be $2y=(j'\!\!+\!j)=2(n\!-\!1)\!-\!(i'\!\!+\!i)$.

Note that for each coupling-power combination $(i'\!,i)$ the
evaluation time for each associated Laurent coefficient is the same,
as they are retrieved simultaneously through \tred's \polythree
construct.

In the following benchmark results, we will not print those that are
identically zero throughout all $\eps$ due to symmetry considerations
(\eg due to color arguments; we make an exception in the following for
$u\bar{u}\rightarrow t\bar{t}$, though, as an example), and in general
we will not give all possible power-coupling combinations. We also
refrain from giving results for every subprocess, although they have
been generated and are available in the process repository, allowing
also for massive initial-state $b$ quarks if applicable.

All results are computed using \oneloop as the scalar provider, since it allows
the option of complex mass arguments, needed if widths are used.

\subsubsection{$p p \rightarrow t \bar{t}$}
\label{sec:ppttbar}

In the case of massless initial-state particles this process consists
of three independent subprocesses: $u \bar{u} \rightarrow t \bar{t}$,
$d \bar{d} \rightarrow t \bar{t}$, and $g g \rightarrow t \bar{t}$,
all tested with the PSP of Table~\ref{tab:psp00tt}. In our process
archive we also provide a distinct initial state for the $b$-quark
channel, in case of a massive $b$ quark.  The fixed-order NLO QCD
corrections to $p p \rightarrow t \bar{t}$ were first calculated
in~\cite{Nason:1987xz,Nason:1989zy,Beenakker:1988bq,Beenakker:1990maa},
while the corresponding NLO EW corrections can be found
in~\cite{Beenakker:1993yr,Kuhn:2005it,Bernreuther:2005is,Kuhn:2006vh,Bernreuther:2006vg,Moretti:2006nf,Bernreuther:2008md,Hollik:2007sw}.

To give an example, for the quark-initiated subprocesses the complete
set of \lo contributions is ($a_0^{(i',i)}\,\widehat{=}\,$
\texttt{tree\_tree g$i'$e$j'$\_g$i$e$j$}, with $j'=n-2-i'$, $j=n-2-i$
and $n=4$)
\begin{itemize}
\item[]
$a_0^{(2,2)}\,\widehat{=}\,$ 
\texttt{tree\_tree g2e0\_g2e0},
$a_0^{(2,0)}\!=\!a_0^{(0,2)}\,\widehat{=}\,$ 
\texttt{tree\_tree g2e0\_g0e2} = \texttt{tree\_tree g0e2\_g2e0},
$a_0^{(0,0)}\,\widehat{=}\,$ \texttt{tree\_tree g0e2\_g0e2},
\end{itemize}
while the complete set of \nlo contributions is ($c_\eps^{(i',i)}\,\widehat{=}\,$ 
\texttt{tree\_loop g$i'$e$j'$\_g$i$e$j$},
with $j'=n-2-i'$, $j=n-i$ and $n=4$)
\begin{itemize}
\item[]
$c_\eps^{(2,4)}\,\widehat{=}\,$ \texttt{tree\_loop g2e0\_g4e0}, 
$c_\eps^{(2,2)}\,\widehat{=}\,$ \texttt{tree\_loop g2e0\_g2e2},
$c_\eps^{(2,0)}\,\widehat{=}\,$ \texttt{tree\_loop g2e0\_g0e4}, 
$c_\eps^{(0,4)}\,\widehat{=}\,$ \texttt{tree\_loop g0e2\_g4e0}, 
$c_\eps^{(0,2)}\,\widehat{=}\,$ \texttt{tree\_loop g0e2\_g2e2},
$c_\eps^{(0,0)}\,\widehat{=}\,$ \texttt{tree\_loop g0e2\_g0e4}. 
\end{itemize}
The following contributions are identically zero throughout all
$\eps$, though, due to color arguments (always producing a color trace
over only one fundamental color matrix): $a_0^{(2,0)}=a_0^{(0,2)}$ and
$c_\eps^{(2,0)}$.  For the gluon-initiated subprocess we only have
$a_0^{(2,2)}$, $c_\eps^{(2,4)}$ and $c_\eps^{(2,2)}$, none of them
being identically zero throughout all $\eps$ (\eg due to color
arguments; the $\order{\alpha_s^2\alpha_e^1}$ contribution has no
double pole, though).  More details in regards to the coupling-power
picture for this process (and also the process $pp\rightarrow jj$; see
next section) are given in the example at the end of
Sec.~\ref{sec:mixedexpansions}.

\begin{table}[H]
\centering

\begin{tabular}{l l|c c}
$u \bar{u} \rightarrow t \bar{t}$ & & \nlox & \recola \\ \hline
$\order{\alpha_s^2\alpha_e^0}$ & $a_0^{(2,2)}$ & 69.83600143751471 & 69.83600143751464 \\
$\order{\alpha_s^1\alpha_e^1}$ & $a_0^{(2,0)}$ & 0 & 0 \\
& $a_0^{(0,2)}$ & 0 & 0 \\
$\order{\alpha_s^0\alpha_e^2}$ & $a_0^{(0,0)}$ & 8.373783006235811 & 8.373783006235795 \\
\hline
$\order{\alpha_s^3\alpha_e^0}$ &$c_2^{(2,4)}$ &
-29.63931955881074 & -29.63931955869816 \\
&$c_1^{(2,4)}$ &
-21.62010557945776 & -21.62010557948297 \\
&$c_0^{(2,4)}$ &
326.7316251339323 & 326.7316251338464 \\
\hline
$\order{\alpha_s^2\alpha_e^1}$ &$c_2^{(2,2)}$ &
-9.879773186270731 & -9.879773186033162 \\
&$c_1^{(2,2)}$ &
100.9206592499375 & 100.9206592501215 \\
&$c_0^{(2,2)}$ &
-264.008143628928 & -264.0081436286176 \\
$\order{\alpha_s^2\alpha_e^1}$ &$c_2^{(0,4)}$ & 
-3.016507999556917e-13 & 3.632294465205632e-11 \\
&$c_1^{(0,4)}$ & 
24.46218467653051 & 24.46218467654894 \\
&$c_0^{(0,4)}$ & 
40.71710556011429 & 40.71710556012410 \\
\hline
$\order{\alpha_s^1\alpha_e^2}$ &$c_2^{(2,0)}$ &
0 & 5.797831350820261e-16 \\
&$c_1^{(2,0)}$ & 0 & 1.627710312214327e-14 \\
&$c_0^{(2,0)}$ & 0 & -7.232455052854060e-15 \\
$\order{\alpha_s^1\alpha_e^2}$ &$c_2^{(0,2)}$ & 
-3.553943887523572 & -3.553943887523602 \\
&$c_1^{(0,2)}$ & 
3.465847464830842 & 3.465847464830841 \\
&$c_0^{(0,2)}$ & 
44.3414894585291 & 44.34148945852903 \\
\hline
$\order{\alpha_s^0\alpha_e^3}$ &$c_2^{(0,0)}$ & 
-1.184647962507978 & -1.184647962463359 \\
&$c_1^{(0,0)}$ & -2.08629577374953 & -2.086295773705025 \\
&$c_0^{(0,0)}$ & -46.41633996223131 & -46.41633996214514 \\
\end{tabular}

\caption{
$u\bar{u}\rightarrow t\bar{t}$. 
The following contributions are identically zero throughout all 
$\eps$, due to color arguments:
$a_0^{(2,0)}=a_0^{(0,2)}$,
$c_\eps^{(2,0)}$.
}

\end{table}

\begin{table}[H]
\centering

\begin{tabular}{l l|c c}
$d \bar{d} \rightarrow t \bar{t}$ & & \nlox & \recola \\ \hline
$\order{\alpha_s^2\alpha_e^0}$ & $a_0^{(2,2)}$ & 69.83600143751471 & 69.83600143751464 \\
$\order{\alpha_s^0\alpha_e^2}$ & $a_0^{(0,0)}$ & 2.807482983131919 & 2.807482983131935 \\
\hline
$\order{\alpha_s^3\alpha_e^0}$ &$c_2^{(2,4)}$ &
-29.63931955881074 & -29.63931955869816 \\
&$c_1^{(2,4)}$ &
-21.62010557945776 & -21.62010557948297 \\
&$c_0^{(2,4)}$ &
326.7316251339323 & 326.7316251338464 \\
\hline
$\order{\alpha_s^2\alpha_e^1}$ &$c_2^{(2,2)}$ &
-2.46994329656692 & -2.469943296594337 \\
&$c_1^{(2,2)}$ &
-10.30104753296668 & -10.30104753304126 \\
&$c_0^{(2,2)}$ &
-876.7537677581068 & -876.7537677580882 \\
$\order{\alpha_s^2\alpha_e^1}$ &$c_2^{(0,4)}$ & 
-5.255491264144508e-14 & 1.550315431586569e-12 \\
&$c_1^{(0,4)}$ & 
2.361140127198419 & 2.361140127198432 \\
&$c_0^{(0,4)}$ & 
3.873233073210399 & 3.873233073210025 \\
\hline
$\order{\alpha_s^1\alpha_e^2}$ &$c_2^{(0,2)}$ & 
-1.191532785098194 & -1.191532785098204 \\
&$c_1^{(0,2)}$ & 
1.161996647440871 & 1.161996647440898 \\
&$c_0^{(0,2)}$ & 
17.55679660782326 & 17.55679660782354 \\
\hline
$\order{\alpha_s^0\alpha_e^3}$ &$c_2^{(0,0)}$ & 
-0.09929439875817378 & -0.09929439875884327 \\
&$c_1^{(0,0)}$ & -4.282220826752567 & -4.282220826748949 \\
&$c_0^{(0,0)}$ & -48.01638812955415 & -48.01638812949420 \\
\end{tabular}

\caption{
$d\bar{d}\rightarrow t\bar{t}$. 
The following contributions are identically zero throughout all 
$\eps$, due to color arguments:
$a_0^{(2,0)}=a_0^{(0,2)}$,
$c_\eps^{(2,0)}$.
}

\end{table}

\begin{table}[H]
\centering

\begin{tabular}{l l|c c}
$g g \rightarrow t \bar{t}$ & & \nlox & \recola \\ \hline
$\order{\alpha_s^2\alpha_e^0}$ & $a_0^{(2,2)}$ & 726.898367145306 & 726.8983671453062 \\
\hline
$\order{\alpha_s^3\alpha_e^0}$ &$c_2^{(2,4)}$ &
-694.1368095396184 & -694.1368095396263 \\
&$c_1^{(2,4)}$ &
-2548.515434412368 & -2548.515434412614 \\
&$c_0^{(2,4)}$ &
-429.2152536196467 & -429.2152536220427 \\
\hline
$\order{\alpha_s^2\alpha_e^1}$ &$c_2^{(2,2)}$ &
0 & 0.000000000000000 \\
&$c_1^{(2,2)}$ &
254.5385946968528 & 254.5385946969382 \\
&$c_0^{(2,2)}$ &
523.671501813473 & 523.6715018106897 \\
\end{tabular}

\caption{
$gg\rightarrow t\bar{t}$. 
}

\end{table}

\subsubsection{$p p \rightarrow j j$}
\label{sec:ppjj}

Fixed-order NLO QCD corrections to $p p \rightarrow jj$ were first
calculated in~\cite{Ellis:1985er}, while the corresponding NLO EW
corrections can be found in~\cite{Moretti:2006ea}.  With electroweak
corrections included, this process contains 37 numerically distinct
subprocesses, including $b$ jets in the final state and $b$ quarks in
the initial state, which are distinct if the $b$ quarks are massive.
This count takes into account symmetries that produce identical
subprocess code. We present a few representative comparisons here for
the entirely massless 2-to-2 PSP in Table~\ref{tab:psp0000}.

More details in regards to the coupling-power picture for this process
(and also the process $pp\rightarrow t\bar{t}$; see previous section) 
are given in the example at the end of 
Sec.~\ref{sec:mixedexpansions}.

\begin{table}[H]
\centering
\begin{tabular}{l l|c c}
$u \bar{u} \rightarrow d \bar{d}$ & & \nlox & \recola \\ \hline
$\order{\alpha_s^2\alpha_e^0}$ & $a_0^{(2,2)}$ & 69.78980394856457 & 69.78980394856457 \\
$\order{\alpha_s^1\alpha_e^1}$ & $a_0^{(2,0)}$ & -0.001242222784080782 & -0.001242222784081074 \\
& $a_0^{(0,2)}$ & -0.001242222784080782 & -0.001242222784081074 \\
\hline
$\order{\alpha_s^3\alpha_e^0}$ &$c_2^{(2,4)}$ &
-59.239425471085 & -59.23942547109301 \\
&$c_1^{(2,4)}$ & 
-175.9794797288048 & -175.9794797287917 \\
&$c_0^{(2,4)}$ &
-98.41451624298804 & -98.41451624298804 \\
\hline
$\order{\alpha_s^2\alpha_e^1}$ &$c_2^{(2,2)}$ &
-12.34049254169433 & -12.34049254170276 \\
&$c_1^{(2,2)}$ &
-66.57794605840587 & -66.57794605839047 \\
&$c_0^{(2,2)}$ &
-641.7235774288396 & -641.7235774288340 \\
$\order{\alpha_s^2\alpha_e^1}$ &$c_2^{(0,4)}$ & 
0.001054431448151187 & 0.001054431447901294 \\
&$c_1^{(0,4)}$ & 
9.889347931105585 & 9.889347931106062 \\
&$c_0^{(0,4)}$ & 
19.48443223966789 & 19.48443223965836 \\\\
%
$d g \rightarrow d g$ & & \nlox & \recola \\ \hline
$\order{\alpha_s^2\alpha_e^0}$ & $a_0^{(2,2)}$ & 25089.40333374167 & 25089.40333374156 \\
\hline
$\order{\alpha_s^3\alpha_e^0}$ &$c_2^{(2,4)}$ &
-34606.88885152432 & -34606.88885153094 \\
&$c_1^{(2,4)}$ & 
-189987.7760195865 & -189987.7760195137 \\
&$c_0^{(2,4)}$ &
-308102.3176212214 & -308102.3176212332 \\
\hline
$\order{\alpha_s^2\alpha_e^1}$ &$c_2^{(2,2)}$ &
-887.3561243980364 & -887.3561243987060 \\
&$c_1^{(2,2)}$ &
-1333.535787238849 & -1333.535787235778 \\
&$c_0^{(2,2)}$ &
-5924.964392741716 & -5924.964392718713 \\\\
%
$g g \rightarrow g g$ & & \nlox & \recola \\ \hline
$\order{\alpha_s^2\alpha_e^0}$ & $a_0^{(2,2)}$ & 44706663.18826243 & 44706663.18826232 \\
\hline
$\order{\alpha_s^3\alpha_e^0}$ &$c_2^{(2,4)}$ &
-85383437.22667754 & -85383437.22669885 \\
&$c_1^{(2,4)}$ & 
-369364175.5404882 & -369364175.5404322 \\
&$c_0^{(2,4)}$ &
-281755147.8249173 & -281755147.8249879 \\
\hline
\end{tabular}
\end{table}

\subsubsection{$p p \rightarrow Z b$}
\label{sec:ppZb}

This process was examined by the authors using \nlox
recently~\cite{Figueroa:2018chn,Figueroa:2018aqv}, where particular
attention was paid to EW corrections and using a massive $b$ quark.
The corresponding fixed-order NLO QCD corrections were first
calculated in~\cite{Campbell:2003dd}.

This process proceeds through only one channel, $bg \rightarrow Zb$.
We present the case of a massive $b$, with the PSP given in
Tab.~\ref{tab:pspb0Zb}.

\begin{table}[H]
\centering
\begin{tabular}{l l|c c}
$b g \rightarrow Z b$ & & \nlox & \recola \\ \hline
$\order{\alpha_s^1\alpha_e^1}$ & $a_0^{(1,1)}$ & 56.70705366755421 & 56.70705366755446 \\
\hline
$\order{\alpha_s^2\alpha_e^1}$ &$c_2^{(1,3)}$ &
-27.07562369811053 & -27.07562369824791 \\
&$c_1^{(1,3)}$ & 
219.1031319518903 & 219.1031319535024 \\
&$c_0^{(1,3)}$ &
1719.562951862683 & 1719.562951868557 \\
\hline
$\order{\alpha_s^1\alpha_e^2}$ &$c_2^{(1,1)}$ &
0 & 0.000000000000000 \\
&$c_1^{(1,1)}$ &
-29.9292665465376 & -29.92926654654536 \\
&$c_0^{(1,1)}$ &
-1381.341046035739 & -1381.341046280566 \\
\end{tabular}
\end{table}

\subsubsection{$p p \rightarrow e^+ e^-$}
\label{sec:ppepem}

To provide a simple test of the complex mass scheme, we examine the
Drell-Yan process mediated both by a photon, and, in contrast to other
tests in this section, by a $Z$ with its width. For this process only,
the parameters $m_Z = 91.1534806191828$ and
$\Gamma_Z = 2.49426637877282$ (GeV) are used, with all other
parameters as in Tab.~\ref{tab:params}. In the following we give
results for the $u\bar{u}\rightarrow e^+e^-$ channel, using the
PSP of Tab.~\ref{tab:psp0000}. 
Fixed-order NLO QCD corrections to the Drell-Yan process were first
calculated in~\cite{Altarelli:1979ub}, while NLO EW
corrections for the neutral current Drell-Yan process can be found in~\cite{Baur:2001ze,Baur:1997wa}.

\begin{table}[H]
\centering
\begin{tabular}{l l|c c}
$u \bar{u} \rightarrow e^+ e^-$ & & \nlox & \recola \\ \hline
$\order{\alpha_s^0\alpha_e^2}$ & $a_0^{(0,0)}$ & 117.4358832604318 & 117.4358832604319 \\
\hline
$\order{\alpha_s^1\alpha_e^2}$ &$c_2^{(0,2)}$ &
-49.84133684602794 & -49.84133684602804 \\
&$c_1^{(0,2)}$ & 
-74.76200526904191 & -74.76200526904215 \\
&$c_0^{(0,2)}$ &
46.59179136175548 & 46.59179136175552 \\
\hline
$\order{\alpha_s^0\alpha_e^3}$ &$c_2^{(0,0)}$ &
-53.99478158319884 & -53.99478158322654 \\
&$c_1^{(0,0)}$ &
53.74087481152656 & 53.74087481148945 \\
&$c_0^{(0,0)}$ &
-2951.1335231532 & -2951.133523154139 \\
\end{tabular}
\end{table}

\subsubsection{$e^+ e^- \rightarrow h t \bar{t}$}
\label{sec:eehtt}

As an example of a lepton-initiated process, we present
$e^+ e^- \rightarrow h t \bar{t}$. It uses the PSP given in 
Tab.~\ref{tab:psp00htt}.
Fixed-order NLO QCD corrections for this process have been 
calculated in~\cite{Dawson:1998ej,Dittmaier:1998dz}, while the corresponding NLO EW
corrections have been presented in~\cite{Denner:2003ri,Denner:2003zp,You:2003zq}.

\begin{table}[H]
\centering
\begin{tabular}{l l|c c}
$e^+ e^- \rightarrow h t \bar{t}$ & & \nlox & \recola \\ \hline
$\order{\alpha_s^0\alpha_e^3}$ & $a_0^{(0,0)}$ & 0.1650875547906978 & 0.1650875547906976 \\
\hline
$\order{\alpha_s^1\alpha_e^3}$ &$c_2^{(0,2)}$ &
0 & 0.000000000000000 \\
&$c_1^{(0,2)}$ & 
0.04539059879996656 & 0.04539059879996588 \\
&$c_0^{(0,2)}$ &
0.4149530650547262 & 0.4149530650547257 \\
\hline
$\order{\alpha_s^0\alpha_e^4}$ &$c_2^{(0,0)}$ &
-0.05254900077578735 & -0.05254900077578772 \\
&$c_1^{(0,0)}$ &
-0.4184851162495272 & -0.4184851162495317 \\
&$c_0^{(0,0)}$ &
-5.119506322697255 & -5.119506321326591 \\
\end{tabular}
\end{table}

\subsubsection{$p p \rightarrow Z t \bar{t}$}
\label{sec:ppZtt}

This process has three (four) distinct subprocesses if the initial state $b$ is
treated as massless (massive). In the following we give benchmark
results for the $u\bar{u}\rightarrow Zt\bar{t}$ and $gg\rightarrow Zt\bar{t}$
channels, using the PSP of Tab.~\ref{tab:psp00Ztt}. 
Fixed-order NLO QCD corrections for this process have been 
calculated in~\cite{Lazopoulos:2008de,Garzelli:2011is,Kardos:2011na,Garzelli:2012bn,Maltoni:2015ena}, while the corresponding NLO EW
corrections have been presented in~\cite{Frixione:2015zaa}.

\begin{table}[H]
\centering
\begin{tabular}{l l|c c}
$u \bar{u} \rightarrow Z t \bar{t}$ & & \nlox & \recola \\ \hline
$\order{\alpha_s^2\alpha_e^1}$ & $a_0^{(2,2)}$ & 0.02418281962609314 & 0.02418281962609323 \\
\hline
$\order{\alpha_s^3\alpha_e^1}$ &$c_2^{(2,4)}$ &
-0.01026350741704608 & -0.01026350741704629 \\
&$c_1^{(2,4)}$ & 
-0.01772187646503733 & -0.01772187646503737 \\
&$c_0^{(2,4)}$ &
0.1034156002235128 & 0.1034156002235174 \\
\hline
$\order{\alpha_s^2\alpha_e^2}$ &$c_2^{(2,2)}$ &
-0.003421169139015138 & -0.003421169139015495 \\
&$c_1^{(2,2)}$ &
-0.02453440885384433 & -0.02453440885384411 \\
&$c_0^{(2,2)}$ &
-0.3832664778286854 & -0.3832664783495138 \\
$\order{\alpha_s^2\alpha_e^2}$ &$c_2^{(0,4)}$ & 
1.834978622567579e-16 & 6.418476861114186e-17 \\
&$c_1^{(0,4)}$ & 
-0.003287144564297587 & -0.003287144564297309 \\
&$c_0^{(0,4)}$ & 
-0.006365140215048126 & -0.006365140215044396 \\\\
%
$g g \rightarrow Z t \bar{t}$ & & \nlox & \recola \\ \hline
$\order{\alpha_s^2\alpha_e^1}$ & $a_0^{(2,2)}$ & 0.01377970228330338 & 0.01377970228330350 \\
\hline
$\order{\alpha_s^3\alpha_e^1}$ &$c_2^{(2,4)}$ &
-0.0131586463963351 & -0.01315864639633449 \\
&$c_1^{(2,4)}$ & 
-0.0189899781008642 & -0.01898997810086434 \\
&$c_0^{(2,4)}$ &
0.05073427513231134 & 0.05073427513231166 \\
\hline
$\order{\alpha_s^2\alpha_e^2}$ &$c_2^{(2,2)}$ &
0 & 0.000000000000000 \\
&$c_1^{(2,2)}$ & 
-0.007929673919194006 & -0.007929673919193275 \\
&$c_0^{(2,2)}$ &
-0.162870316537v5728 & -0.1628703165354349 \\
\end{tabular}
\end{table}

\subsubsection{$p p \rightarrow W^+ t \bar{t}$}
\label{sec:ppWtt}

This process proceeds through only one distinct channel. The following
results are obtained using the PSP in Table~\ref{tab:psp00Wtt}.
Fixed-order NLO QCD corrections for this process have been 
calculated in~\cite{Campbell:2012dh,Garzelli:2012bn,Maltoni:2015ena}, while the corresponding NLO EW
corrections have been presented in~\cite{Frixione:2015zaa,Frederix:2017wme}.

\begin{table}[H]
\centering
\begin{tabular}{l l|c c}
$u \bar{d} \rightarrow W^+ t \bar{t}$ & & \nlox & \recola \\ \hline
$\order{\alpha_s^2\alpha_e^1}$ & $a_0^{(2,2)}$ & 0.03893217777719004 & 0.03893217777718989 \\
\hline
$\order{\alpha_s^3\alpha_e^1}$ &$c_2^{(2,4)}$ &
-0.01652332943619126 & -0.01652332943619256 \\
&$c_1^{(2,4)}$ & 
-0.02847293904026631 & -0.02847293904026624 \\
&$c_0^{(2,4)}$ &
0.2071974285545872 & 0.2071974285546108 \\
\hline
$\order{\alpha_s^2\alpha_e^2}$ &$c_2^{(2,2)}$ &
-0.003442360299206428 & -0.003442360299207081 \\
&$c_1^{(2,2)}$ &
-0.01496796182114426 & -0.01496796182114191 \\
&$c_0^{(2,2)}$ &
-0.8087579238035377 & -0.8087579204322376 \\
$\order{\alpha_s^2\alpha_e^2}$ &$c_2^{(0,4)}$ & 
2.270873927545318e-15 & -3.408731630294426e-16 \\
&$c_1^{(0,4)}$ & 
0.001082295240220039 & 0.001082295240219950 \\
&$c_0^{(0,4)}$ & 
-0.003917997147393862 & -0.003917997147326103 \\
\end{tabular}
\end{table}

\subsubsection{$p p \rightarrow e^+ \nu_e \mu^- \bar{\nu}_{\mu}$}
\label{sec:ppWW}

This process is intended to reflect $WW$ production with decays, and
includes all appropriate intermediate particles for this final
state. It only has quark-initiated channels due to the purely EW final
state. The EW corrections to
$u \bar{u} \rightarrow e^+ \nu_e \mu^- \bar{\nu}_{\mu}$, in the
complex mass scheme with $W$ and $Z$ widths, first calculated
in~\cite{Kallweit:2017khh}, were recently tested among various codes
in \cite{Bendavid:2018nar}. The parameters $m_Z = 91.1534806191828$,
$\Gamma_Z = 2.49426637877282$, $m_W = 80.3579736098775$, and
$\Gamma_W = 2.08429899827822$ are used instead of the masses of
Tab.~\ref{tab:params}, with all other parameters the same. The
subprocesses below use the PSP of Tab.~\ref{tab:psp000000}.

\begin{table}[H]
\centering
\begin{tabular}{l l|c c}
$u \bar{u} \rightarrow e^+ \nu_e \mu^- \bar{\nu}_{\mu}$ & & \nlox & \recola \\ \hline
$\order{\alpha_s^0\alpha_e^4}$ & $a_0^{(0,0)}$ & 7.887069176044188e-05 & 7.887069176044208e-05 \\
\hline
$\order{\alpha_s^1\alpha_e^4}$ &$c_2^{(0,2)}$ &
-3.347376122333766e-05 & -3.347376122333511e-05 \\
&$c_1^{(0,2)}$ & 
-5.021064183500852e-05 & -5.021064183503748e-05 \\
&$c_0^{(0,2)}$ &
7.08451416208973e-05 & 7.084514162084351e-05 \\
\hline
$\order{\alpha_s^0\alpha_e^5}$ &$c_2^{(0,0)}$ &
-3.626324132527428e-05 & -3.626324132527568e-05 \\
&$c_1^{(0,0)}$ &
-0.0002743890608279887 & -0.0002743890608279680 \\
&$c_0^{(0,0)}$ &
-0.005521968082280729 & -0.005521968082281660 \\\\
$d \bar{d} \rightarrow e^+ \nu_e \mu^- \bar{\nu}_{\mu}$ & & \nlox & \recola \\ \hline
$\order{\alpha_s^0\alpha_e^4}$ & $a_0^{(0,0)}$ & 0.000510020443960627 & 0.0005100204439606252 \\
\hline
$\order{\alpha_s^1\alpha_e^4}$ &$c_2^{(0,2)}$ &
-0.0002164593992913518 & -0.0002164593992913509 \\
&$c_1^{(0,2)}$ & 
-0.0003246890989370246 & -0.0003246890989370279 \\
&$c_0^{(0,2)}$ &
0.0004289020803322432 & 0.0004289020803322372 \\
\hline
$\order{\alpha_s^0\alpha_e^5}$ &$c_2^{(0,0)}$ &
-0.0001803828327427987 & -0.0001803828327427868 \\
&$c_1^{(0,0)}$ &
-0.00229568929769725 & -0.002295689297697336 \\
&$c_0^{(0,0)}$ &
-0.02911547502773986 & -0.02911547502774042 \\
\end{tabular}
\end{table}

\subsubsection{$p p \rightarrow e^+ e^- \mu^+ \mu^-$}
\label{sec:ppZZ}

This process is similar to the previous one~\ref{sec:ppWW}, reflecting
$ZZ$ production with decays, and uses the same masses, widths, and
PSP. NLO EW corrections were first calculated
in~\cite{Biedermann:2016lvg}, and the corresponding one-loop matrix
elements were cross checked among several
automated OLPs in \cite{Bendavid:2018nar}.

\begin{table}[H]
\centering
\begin{tabular}{l l|c c}
$u \bar{u} \rightarrow e^+ e^- \mu^+ \mu^-$ & & \nlox & \recola \\ \hline
$\order{\alpha_s^0\alpha_e^4}$ & $a_0^{(0,0)}$ & 0.000236139465881233 & 0.0002361394658812330 \\
\hline
$\order{\alpha_s^1\alpha_e^4}$ &$c_2^{(0,2)}$ &
-0.0001002207020108752 & -0.0001002207020108742 \\
&$c_1^{(0,2)}$ & 
-0.0001503310530163127 & -0.0001503310530163135 \\
&$c_0^{(0,2)}$ &
0.0001734208724655893 & 0.0001734208724656161 \\
\hline
$\order{\alpha_s^0\alpha_e^5}$ &$c_2^{(0,0)}$ &
-0.0001837379536866157 & -0.0001837379536866150 \\
&$c_1^{(0,0)}$ &
-0.001085519471500961 & -0.001085519471501003 \\
&$c_0^{(0,0)}$ &
-0.01202757057543053 & -0.01202757057542850 \\\\
$d \bar{d} \rightarrow e^+ e^- \mu^+ \mu^-$ & & \nlox & \recola \\ \hline
$\order{\alpha_s^0\alpha_e^4}$ & $a_0^{(0,0)}$ & 2.094078745814468e-05 & 2.094078745814468e-05 \\
\hline
$\order{\alpha_s^1\alpha_e^4}$ &$c_2^{(0,2)}$ &
-8.887546229867955e-06 & -8.887546229867840e-06 \\
&$c_1^{(0,2)}$ & 
-1.333131934480209e-05 & -1.333131934480482e-05 \\
&$c_0^{(0,2)}$ &
1.024691111771567e-05 & 1.024691111771076e-05 \\
\hline
$\order{\alpha_s^0\alpha_e^5}$ &$c_2^{(0,0)}$ &
-1.407194819729486e-05 & -1.407194819729254e-05 \\
&$c_1^{(0,0)}$ &
-0.0001170110505691798 & -0.0001170110505691632 \\
&$c_0^{(0,0)}$ &
-0.0006080499354058187 & -0.0006080499354053612 \\
\end{tabular}
\end{table}

\subsection{\nlox Timings}

For the tested processes of Sec.~\ref{subsec:processes}, we also present average
PSP timings for each loop contribution $c_\eps^{(i',i)}$,
which can be run independently in \nlox. As the entire polynomial in
$\epsilon$ is returned simultaneously as a \polythree, there is only one time
for all $c_\eps^{(i',i)}$ terms for a given $(i', i)$.

These timings were tested by interfacing \nlox with a simple PSP
generator. Since \nlox sometimes evaluates at higher precision, for a better
representative time we allow the generator to sample much 
of phase space, cutting out only singular regions, using as a guide 
$p_T$ and $\Delta R$ cuts similar to those found in LHC analyses, 
for energies similar to typical LHC partonic energies. We generate 1000
points, and typically 500 to 900 points pass these cuts,
depending mostly on particle multiplicity. The final average time number
presented for each subprocess is the total run time
(after program initialization) divided by the number of points that pass the
cuts (those that don't pass are not evaluated). As in the PSP checks, \oneloop
is used, though the runtime does not strongly depend on the scalar provider
as evaluation of scalar integrals is not a large contributor
to the overall runtime, especially for large processes.

The timing benchmarks are meant to serve as a 
useful estimate of the efficiency of \nlox. As benchmarks are 
typically very machine and environment dependent, we refrain from 
direct timing comparisons to other codes, but we generally 
find very good performance of \nlox.

\begin{table}
\begin{centering}
\begin{tabular}{l | c c c}
& & $t_{avg}$ [ms] & \\
\raisebox{2.5ex}{Subprocess} & $c_{\eps}^{(max,max)}$ & $c_{\eps}^{(max,max-2)}$ & $c_{\eps}^{(max-2,max)}$ \\
\hline\\[-1.5ex]
$u \bar{u} \rightarrow t \bar{t}$ & 0.22 & 0.48 & 0.22 \\
$d \bar{d} \rightarrow t \bar{t}$ & 0.23 & 0.47 & 0.19 \\
$g g \rightarrow t \bar{t}$ & 0.74 & 1.5 & - \\
$u \bar{u} \rightarrow d \bar{d}$ & 0.19 & 0.48 & 0.20 \\
$d g \rightarrow d g$ & 0.45 & 0.68 & - \\
$g g \rightarrow g g$ & 1.6 & - & - \\
$b g \rightarrow Z b$ & 0.61 & 1.9 & - \\
$u \bar{u} \rightarrow e^+ e^-$ & 0.11 & 0.92 & - \\
$e^+ e^- \rightarrow h t \bar{t}$ & 2.1 & 39 & - \\
$u \bar{u} \rightarrow Z t \bar{t}$ & 4.3 & 17 & 5.2 \\
$g g \rightarrow Z t \bar{t}$ & 32 & 69 & - \\
$u \bar{u} \rightarrow e^+ \nu_e \mu^- \bar{\nu}_{\mu}$ & 6.3 & 237 & - \\
$u \bar{u} \rightarrow e^+ e^- \mu^+ \mu^-$ & 24 & 1116 & - \\

\end{tabular}
\caption{Timings for various contributions to processes in this version of
\nlox, in ms. $c_\eps^{(max, max)}$ denotes the set of one-loop Laurent
coefficients with the highest possible power of $g_s$ for the subprocess in
question. Combinations that produce a trivial zero are represented with "-".
These benchmarks were performed on an Intel i7 950 (3.07 GHz) running 
Scientific Linux 7.3, compiled with gcc 4.8.5 with option -Og.}
\end{centering}
\end{table}

\section{Summary and Outlook}
\label{sec:summary}

We have presented the \nlox package, a one-loop provider for the
calculation of NLO QCD and EW corrections to SM processes with up to
six external particles. Based on a traditional Feynman-diagram
approach, \nlox optimizes parsing and storing of recurrent building
blocks, as realized at various stages during \cpp process code
generation and in particular by the \tred library.

\tred accumulates and stores a list of all tensor coefficients which
appear during the recursive tensor-integral reduction, along with
their dependencies. It is designed to handle multiple reduction
methods simultaneously, deal with additional coefficients at runtime,
and uses an efficient numerical approach with internal stability
checks, only calculating what is requested and needed. It is
extensible to new reduction methods, and we plan to add more in the
future. It can be used as a standalone library, and we may issue
standalone releases with updates in the future.

We have reviewed all the information necessary to understand the
pre-generated processes that are released and the underlying code
functionalities. More pre-generated processes will be added to the
repository as they become available or by request. We will follow with
a release of the source code for process code generation. Alongside,
further developments of the code will be documented and released as
soon as they are implemented and tested. In particular, in the short
term we expect to improve the OLP interface of \nlox, to add the
capability of computing polarized matrix elements as well as color-
and spin-correlated matrix elements, to test the numerical accuracy of
the results using a more complete set of cross checks, and to allow
for the computation of processes with more than six external legs. In
the long term, we plan to allow for a more flexible and more
transparent user implementation of different models.

\section{Acknowledgments}
We would like to thank T.~Schutzmeier for the initial working
design of \nlox, and  D.~Wackeroth for sharing her expertise and
knowledge of QCD and EW one-loop calculations.  This work has
been supported by the U.S. Department of Energy under grant DE-SC0010102.
C.R. acknowledges current support by the European Union’s Horizon 2020
research and innovative programme, under grant agreement No. 668679.
S.H., L.R., and C.R. are grateful for the hospitality of the Kavli
Institute for Theoretical Physics (KITP) during the workshop on
\textit{LHC Run II and the Precision Frontier} where part of this work
was being developed.  Their research at the KITP was supported in part
by the National Science Foundation under Grant No. NSF PHY-1748958.
L.~R. would like to also thank the Aspen Center for Physics for the
hospitality offered while parts of this work were being completed.



\appendix

\section{Details on Renormalization}
\label{app:renormalization}

We briefly collect in this appendix some general aspects of the \nlox
renormalization procedure that may be of interest to the
user.

\subsection{Counterterms}
\label{sec:cts}

\nlox's approach to renormalization is in terms of counterterm (CT)
diagrams. In the \nlox model a set of CT Feynman rules is specified.
Upon generation of diagrams, also all possible CT diagrams are
generated, and sorted by coupling powers and associated to the
corresponding set of one-loop diagrams.
There are two types of corrections, \ie  QCD and
 EW 
corrections, divided in three cases each, \ie corrections
 to QCD 
vertices,  to EW vertices and to propagators. The CT Feynman 
rules are organized accordingly (we suppress possible color indices;
global minus signs and factors of $i$ are accounted for at the end of 
the calculation):

\begin{enumerate}
\item[$\bullet$] QCD corrections.
\item[-] To QCD vertices:
  $g_s^2\big(\sum_ic_i\delta Z_i\big)V^{(0)}$, where
  $i=\{g_s\},\{G,q,Q\}$. For $i=\{g_s\}$ we have $c_{g_s}=1$ for each
  factor of $g_s$ that appears in the corresponding tree vertex
  $V^{(0)}$ (which already includes the corresponding factors of $g_s$ in its definition), while for $i=\{G,q,Q\}$ we have
  $c_{i}=\frac{1}{2}$ for each strongly charged field of the type of
  $i$ that appears in the corresponding $V^{(0)}$.
\item[-] To EW vertices: $g_s^2\big(\sum_ic_i\delta Z_i\big)V^{(0)}$,
  where $i=\{q,Q\},\{m_Q\}$. For $i=\{q,Q\}$ we have
  $c_{i}=\frac{1}{2}$ for each electroweakly charged field of the type
  of $i$ that appears in the corresponding tree vertex $V^{(0)}$
  (which already includes the corresponding factors of
  $g_e$ in its definition), while for $i=\{m_Q\}$ we have $c_{m_Q}=1$
  for each factor of $m_Q$ that appears in the corresponding
  $V^{(0)}$.
\item[-] The propagator CT diagrams are 
  implemented via standard insertions: $g_s^2\big(p\sla\delta Z_Q
  -m_Q(\delta Z_Q+\delta Z_{m_Q})\big)$ for quarks $Q$ with mass $m_Q$
  (which can also be massless quarks $q$, in which case $m_Q=m_q=0$),
  and $g_s^2(p_\mu p_\nu-p^2g_{\mu\nu})\delta Z_G$ for gluons, where
  $p$ denotes the four-momentum flowing through the propagator.
\end{enumerate}

\begin{enumerate}
\item[$\bullet$] EW corrections.
\item[-] To EW vertices: $g_e^2\big(\sum_ic_i\delta
  Z_i\big)V^{(0)}$, where for each sum over
  the $\delta Z_i$, depending on the underlying corresponding tree 
  vertex $V^{(0)}$ (which already includes the corresponding
  tree factors of $g_e$ in its
  definition), we follow the
  prescription as given in Appendix~A in
  Ref.~\cite{Denner:1991kt} (for a mass
  renormalization constant for a mass $m$ we use the notation 
  $\delta Z_m = \delta m/m$).
\item[-] To QCD vertices: As in the preceding case, 
  with the difference that only $q\bar{q}G$ or
  $Q\bar{Q}G$ tree vertices are considered (which already include the
  corresponding factors of $g_s$ in their definition), which also means that no
  $\delta Z_i$ for the EW coupling, the masses or the weak mixing angle
  are considered.
\item[-] The propagator CT diagrams are 
  implemented via standard insertions, where we follow the prescription 
  as given in Appendix~A in
  Ref.~\cite{Denner:1991kt}, factoring out the corresponding coupling power 
  $g_e^2$ already at the level of the CT Feynman rules for the
  propagators.
\end{enumerate}

In the above description $\delta Z_i$ denote the $i$ renormalization
constant, where $i$ can denote either a field or a mass or a coupling
(we specify them further down below).  In addition there is external
field renormalization: For each external field $X$ in a given
subprocess and each tree diagram $D_i^{(0)}$, we consider a CT diagram
$\frac{1}{2}\delta\bar{R}_XD_i^{(0)}$. As only our QCD renormalization
has non-trivial $\delta\bar{R}_X$, this results in the total external
field CT for that given subprocess to be of the form
$\sum_{\{X\}}\frac{1}{2}\delta\bar{R}_X\sum_iD_i^{(0)}
=\sum_{\{X\}}\frac{1}{2}\delta\bar{R}_XA^{(0)}$,
where $A^{(0)}$ is the corresponding tree amplitude to the tree
diagrams $D_i^{(0)}$.

\subsection{EW Renormalization}
\label{sec:ewrenorm}

In regards to EW renormalization, \ie the implementation of the
renormalization constants in terms of the EW self energies, and the
implementation of the EW self energies, we follow
Ref.~\cite{Denner:1991kt}. However, Ref.~\cite{Denner:1991kt} uses a
small photon mass $\lambda$ to regulate IR singularities, whereas we
use dimensional regularization.\footnote{In our code we implement the
  self energies that enter the EW renormalization constants without
  the common factor of $-\alpha_e/(4\pi)$, as we account for a factor
  of $g_e^2$ in the CT Feynman rules (see \ref{sec:cts}), for the
  purpose of coupling-power counting, and for the rest globally at the
  end of the calculation.}  This changes the pole and finite parts of
the photon terms in the fermion self energies for massless
fermions. With respect to the expressions of $\Sigma_{ij}^{f,L/R/S}$
in App.~B of Ref.~\cite{Denner:1991kt} we thus have slightly different
expressions for the respective photon terms, in the cases $m_{f,i}=0$,
making the replacements
$[2B_1(p^2,m_{f,i}=0,\lambda)+1]\rightarrow2B_1(p^2,m_{f,i}=0,0)$ and
$[4B_0(p^2,m_{f,i}=0,\lambda)-2]\rightarrow4B_0(p^2,m_{f,i}=0,0)$ in
these cases. For non-zero $m_{f,i}$ we recover the same expressions
for $\Sigma_{ij}^{f,L/R/S}$ as in App.~B of Ref.~\cite{Denner:1991kt},
but with $\lambda=0$.  \color{black}

As EW input scheme choices \nlox provides both the $\alpha(0)$ and the
$G_\mu$ EW input schemes~\cite{Hollik:1988ii,Denner:1991kt,
  Dittmaier:2001ay, Andersen:2014efa}. We remind that, according to
the $G_\mu$ input scheme, the Born-level coupling factors
$\alpha_e(0)$ are
replaced by $\alpha_{G_\mu}$:
\begin{equation}
\label{eq:alphagmu}
\alpha_e(0) \to \alpha_{G_\mu}=\frac{\sqrt{2}G_\mu M_W^2}{\pi} 
\left(1-\frac{M_W^2}{M_Z^2}\right)\,\,\,,
\end{equation}
changing the set of independent parameters from
$\{\alpha_e(0),M_W,M_Z\}$ to $\{G_\mu,M_W,M_Z\}$, and the coupling
renormalization constant $\delta Z_e$ receives a contribution from
$\Delta r$~\cite{Sirlin:1980nh}, which describes the EW one-loop
corrections to muon decay.  Consequently, a generic
$O(\alpha_s^x\alpha_e^y)$ term in the expansion of a one-loop
amplitude squared in the $G_\mu$-scheme is related to the
corresponding term in the $\alpha(0)$-scheme as follows:
\begin{equation}
\label{eq:gmu}
{\cal O}(\alpha_s^x\alpha_e^y)_{G_\mu}=
\left[{\cal O}(\alpha_s^x\alpha_e^y)_{\alpha(0)}-
\Delta r \, \alpha_e(0) {\cal O}(\alpha_s^x\alpha_e^{y-1})\right] \frac{\alpha_{G_\mu}}{\alpha_e(0)}\,,
\end{equation}
where $O(\alpha_s^x\alpha^{y-1})$ denotes the corresponding term in
the expansion of the amplitude squared at one order less in $\alpha$.
Per default the $\alpha(0)$ EW input scheme is used and the
corresponding results in the $G_\mu$ scheme are obtained via
Eq.~(\ref{eq:gmu}).  Apart from replacing $\alpha_e(0)$ by
$\alpha_{G_\mu}$ using Eq.~(\ref{eq:alphagmu}), switching to the
$G_\mu$ scheme technically simply consists in replacing the charge
renormalization constant $\delta Z_e$ with $\delta Z_e-\Delta r/2$.

The implementation of the complex-mass scheme follows the description
in Ref.~\cite{Denner:2005fg}, where the choice in \nlox is to follow
the approximation that expands self-energies with complex squared
momenta around real squared momenta, as also described in
Ref.~\cite{Denner:2005fg}.

\subsection{QCD Renormalization}
\label{sec:qcdrenorm}

In the following we denote the 't\,Hooft-mass in dimensional
regularization by $\mu$.  Factors of
$S_\eps\alpha_s/(4\pi)$, with $S_\eps=(4\pi)^\eps/\Gamma(1-\eps)$, are
commonly factored out.  $S_\eps/\eps=\Delta+\order{\eps}$, where
$\Delta=1/\eps-\gamma_E+\log(4\pi)$ denotes the $\msbar$ pole. If we
want to be specific about the $\msbar$ UV pole, then we write
$S_\eps/\eps_\uv$. Similarly, if we want to be specific about the
$\msbar$ IR pole, then we write $S_\eps/\eps_\ir$.  Factors of
$(m^2/\mu^2)^{-\eps}$, where $m$ is a mass, are not commonly factored
out, but are rather expanded in $\eps$.  Furthermore we have $C_A=N_c$
and $C_F=(N_c^2-1)/(2N_c)$, as well as $T_R=1/2$. In numerical
calculations we use $N_c=3$.

We renormalize the mass and wave function of a massive quark $Q$ of
non-zero mass $m_Q$ in a modified on-shell scheme, where the
corresponding renormalization constants only contain UV poles, \ie
\footnote{ In our code we implement the QCD renormalization constants
  without the common factor of $-\alpha_s/(4\pi)$, as we account for a
  factor of $g_s^2$ in the CT Feynman rules (see \ref{sec:cts}), for
  the purpose of coupling-power counting, and for the rest globally at
  the end of the calculation.}
\begin{align}
\label{eq:dZmQnlox}
\delta Z_{m_Q}
&=
-\frac{\alpha_s}{4\pi}\,C_FS_\eps
\bigg(\frac{3}{\eps_\uv}-3\ln\Big(\frac{m_Q^2}{\mu^2}\Big)+4\bigg)
\,,\\
\label{Eq:dZQnlox}
\delta Z_Q
&=
-\frac{\alpha_s}{4\pi}\,C_FS_\eps
\Big(\frac{1}{\eps_\uv}-3\ln\bigg(\frac{m_Q^2}{\mu^2}\Big)+4\bigg)
\,.
\end{align}

We renormalize the wave function of a massless quark $Q=q$ of mass
$m_Q=m_q=0$ in the $\msbar$ scheme, \ie we use
\begin{align}
\delta Z_q
=
-\frac{\alpha_s}{4\pi}\,C_FS_\eps\frac{1}{\eps_{\uv}}\,.
\end{align}

The gluon wave function is also renormalized in the $\msbar$ scheme,
except for the heavy-quark contributions which are subtracted at zero
momentum~\cite{Collins:1978wz,Nason:1987xz}, such that
\begin{align}
\label{eq:dZGnlox}
\delta Z_G(\{q,Q\})
=
-\frac{\alpha_s}{4\pi}\,2S_\eps
\bigg(
\,\frac{1}{\eps_\uv}\,
\Big(
\frac{2T_R(n_{lf}\!+\!n_{hf})}{3}
-\frac{5}{2}\frac{C_A}{3}
\Big)
\,-\,\frac{2T_R}{3}\!\!\!\sum_{f\in\{Q\}}\!\!\!\ln\Big(\frac{m_f^2}{\mu^2}\Big)
\,\bigg)
\,,
\end{align}
where $n_{lf}$ and $n_{hf}$ denote the number of light and heavy quark
flavors respectively (which are the respective dimensions of the
corresponding sets $\{q\}$ and $\{Q\}$ above). Consequently, the
coupling renormalization constant is
\begin{align}
\label{eq:dZgsnlox}
\delta Z_{g_s}(\{q,Q\})
=
-\frac{\alpha_s}{4\pi}\,S_\eps
\bigg(
\,\frac{1}{\eps_\uv}\,
\Big(
\frac{11}{2}\frac{C_A}{3}
-\frac{2T_R(n_{lf}\!+\!n_{hf})}{3}
\Big)
\,+\,\frac{2T_R}{3}\!\!\!\sum_{f\in\{Q\}}\!\!\!\ln\Big(\frac{m_f^2}{\mu^2}\Big)
\,\bigg)\,.
\end{align}

For each external field $X$ in a given process we consider a term
$(1/2)\delta\bar{R}_X=(1/2)\big(\bar{R}_X-1\big)$ times the
corresponding Born contribution, where $\bar{R}_X=R_X/Z_X$ denotes the
renormalized residue.  For massive and massless quarks we use
\begin{align}
\label{eq:dRbarQnlox}
\delta\bar{R}_Q
&=
-\frac{\alpha_s}{4\pi}\,C_FS_\eps\frac{2}{\eps_\ir}
\,,\\
\label{eq:dRbarqnlox}
\delta\bar{R}_q
&=
-\frac{\alpha_s}{4\pi}\,C_FS_\eps\frac{(-1)}{\eps_{\ir}}
\,,
\end{align}
respectively, whereas for gluons we use
\begin{align}
\label{eq:dRbarGnlox}
\delta\bar{R}_G(\{q',Q'\},\{q,Q\})
&=
-\frac{\alpha_s}{4\pi}\,2S_\eps
\bigg(
\,\frac{(-1)}{\eps_\ir}\,
\Big(
\frac{2T_R\,n_{lf}'}{3}
-\frac{5}{2}\frac{C_A}{3}
\Big)
\,-\,\frac{2T_R}{3}\!\!\!\!\!\!\!\!\!\!
\sum_{\scriptstyle f\in\{Q'\}\setminus\{Q\}}\!\!\!\!\!\!\!\!\!\!
\ln\Big(\frac{m_f^2}{\mu^2}\Big)
\,\bigg)\,,
\end{align}
where we introduce two different sets $\{q,Q\}$ and $\{q',Q'\}$, with
$\{q'\}\subseteq\{q\}$ (of dimensions $n'_{lf}\leq n_{lf}$) and
$\{Q'\}\supseteq\{Q\}$ (of dimensions $n'_{hf}\geq n_{hf}$).  In the
commonly used prescription in
Refs. ~\cite{Collins:1978wz,Nason:1987xz} to decouple a given number
of inactive flavors, one usually considers all active flavors in the
set $\{q\}$ to be massless, and to be renormalized in $\msbar$, while
the masses of all of the inactive flavors in the set $\{Q\}$ are
usually considered larger than the typical physical scale of a given
process, and are decoupled by renormalization through zero-momentum
subtraction.  In this context, the active flavors are often referred
to as light flavors, whereas the inactive flavors are referred to as
heavy flavors.  In \nlox the user may also chose to only consider the
first $n'_{lf}$ flavors to be massless, such that there are $n_{lf}$
active and $n_{hf}$ inactive flavors $\{q,Q\}$, but
$n'_{lf}\leq n_{lf}$ massless and massive flavors $\{q',Q'\}$, with
$n'_{lf}+n'_{hf}=n_{lf}+n_{hf}$.  In particular, in cases where one
wants to study the effects of keeping the $n_{lf}$'th flavor massive,
in an $n_{lf}$-flavor scheme, this is of importance (see \eg
Ref.~\cite{Figueroa:2018chn}).  For $\{q'\}=\{q\}$ and $\{Q'\}=\{Q\}$
the commonly used prescription in
Refs.~\cite{Collins:1978wz,Nason:1987xz} is recovered.

\section{Details on Tensor Reduction}

In this appendix we give further details on the technical use and operation
of the \tred library. Header files described in this appendix, as well as the
\tred source, are found in the \texttt{tred} directory of the \nlox package,
whose location can be found in the online documentation on
\url{http://www.hep.fsu.edu/~nlox}.

\subsection{The \tred Class}
\label{sec:tred-class}

The primary interface for the \tred library is the class \tred, which
mangages the storage and computation of all tensor coefficients, and
interfaces to external scalar libraries. Here we describe some of its more
important interface functions. The class and its function declarations
may be found in the file \texttt{tred.h}; some details of function
calls are omitted in the following. All functions described in this section
are methods of the \tred class unless otherwise stated.

\tred stores the names of external particle momenta and masses
symbolically, as the recursion dependencies in the reduction only depend
on the corresponding
symbols, not their actual values. The first step in using the \tred library is to
create a \tred object, given a basis of these momentum and mass
name (which are \cpp strings, stored inside the \tred object for matching
later).
In this way, \tred can determine symbolically which tensor coefficients are
being requested and relate them to others it needs or stores. To
construct a \tred object, one first builds a \texttt{vector} containing
names of momenta used in the process, \eg \texttt{"p1"}. Masses,
being fixed complex numbers, are actually assembled as \cpp \texttt{pair}s of
symbolic strings and their values and assembled into another \texttt{vector}.
These momenta and masses then allow \tred to determine a symbolic
basis for tensor coefficient integrals, and to store and retrieve mass
and momentum values as needed. To create a \tred object from these
\texttt{vectors}, one calls the constructor

\texttt{TRed(momenta, masses)}

\noindent
with \texttt{momenta} and \texttt{masses} the aforementioned assembled
\texttt{vector}s. There are two additional optional arguments to the
constructor. The third takes an integer parameter representing the
number of decimal digits invariants are assumed to be accurate to
relative to other invariants for a given PSP, so that \tred can
internally determine whether these invariants are meant to be
identical. The user may want to adjust this depending on the level of
accuracy expected for the PSP given to \tred. The fourth argument is
an integer representing the allowed methods of the reduction
(see Sec.\ref{subsec:tredfeat}); the default is to use
Passarino-Veltman (PV) for 3- and 4-point
reduction and Denner and Dittmaier's 5/6-point
reduction schemes (E\_DD). The \texttt{method} integer is
assembled bitwise. For example, the default is

\texttt{method = PV\_Reduction | E\_DD\_Reduction}.

The names are aliases to numbers found in \texttt{methods.h}.

The tensor coefficients required are handled and stored in \tred as
\texttt{CoefficientNode} objects. These objects contain
information about the type of coefficient and stores pointers to its
dependencies. They can be used to retrieve the value of the tensor coefficient
depending on the active reduction method through the \texttt{CoefficientNode}
method \texttt{value()}, returning a \polythree object containing the result
for the coefficient, after it has been calculated using the active reduction
method(s).

To add a tensor coefficient, one can call
\texttt{push\_coefficient()}, which takes as arguments the number of
particles comprising the loop integral, a \texttt{vector} of
denominator momenta (strings built from combinations of external
momenta, \eg \texttt{"p1+p2"}), a \texttt{vector} of mass names (\eg
\texttt{"mt"}), and a \texttt{vector} of integers representing the
tensor coefficient indices.  These names should correspond to momenta
and masses from the basis of names given to the \tred constructor when
the object is created; \tred parses the strings from +/- combinations
of external momenta in the denominators automatically.
\texttt{push\_coefficient()} returns a pointer to a
\texttt{CoefficientNode} object. \tred will create the node or find it
already existing as appropriate when \texttt{push\_coefficient()} is
called. Other functions are also available if one does not want to
build a general coefficient, but specify an $A$, $B$, $C$ coefficient,
etc. specifically, \eg \texttt{add\_a\_coeff()}.\footnote{These can be
  more user-friendly for those who want to use the \tred library
  standalone as they have a fixed number of mass and momentum
  arguments and do not require building \cpp \texttt{vector}s, but
  they are not used by the automated process generation package
  \nlox.}  As discussed in the main body of the text, coefficients
will create a separate method-dependent object for each allowed
method, and create the necessary dependencies for each as
needed. Pointers to these are stored in the \texttt{CoefficientNode}
objects for retrieval as needed, but only unique ones are
stored in the \tred object.  Tensor coefficients only need to be
requested from \tred once after the object is built, after which they
and their dependencies persist for the life of the \tred object.

To actually calculate the values of coefficients for a given PSP, the
numerical momenta must be updated. To do this, one must pass a \texttt{vector}
containing the four-momenta to the function \texttt{set\_momenta()}. For this
purpose, a \texttt{FourMomentum}
class has been defined inside \tred, which also contains useful
functions for operating on four-momenta; creating a \texttt{FourMomentum}
object is straightforward: \texttt{FourMomentum(pE, px, py, pz)}, with
the arguments being double precision values of
$E$, $p_x$, $p_y$ and $p_z$, in GeV. To
update the momenta stored in \tred, one assembles a \texttt{vector} of
\texttt{pair}s of
momentum strings and values (\texttt{FourMomentum} objects) as the mass
pairs are created for the \tred constructor, then call \texttt{set\_momenta()}
with the \texttt{vector} as the argument.

For dynamic scale choices, one must update the scale used to compute
the integrals. This is accomplished by the \texttt{set\_musq()}
function.

Finally, all desired coefficients are evaluated with the
\texttt{evaluate()} function, storing values for all coefficients that have
been requested through \texttt{push\_coefficient()} and their dependencies.
This should be done after each PSP is updated, but before one attempts to
retrieve the values through the \texttt{CoefficientNode} \texttt{value()}
function.

\tred also manages the kinematics cache and interfaces to the scalar
integral providers, though the user does not need to know the inner
workings of this. To change the scalar integral provider, see
\texttt{evaluation.h}.

\subsection{The \polythree Class}
\label{sec:poly3}

As discussed in the main body of the text, dimensional regularization
requires an expansion in powers of $\epsilon$, specifically three
powers are needed at one-loop order. Rather than handle the expansion
in $\epsilon$ entirely analytically in the scripts, much of the work
is done numerically through a class, \polythree, dedicated to that
purpose. This has the advantage of keeping expressions organized by
their purpose.

The \polythree class is defined in \tred in the file
\texttt{poly3.h}, though it is used in \nlox for process-dependent
code as well. \polythree objects store the coefficients
$a,b,c$ of a three-term
polynomial for a given
minimum order $n$:
\begin{align}
a \epsilon^n + b \epsilon^{n+1} + c \epsilon^{n+2} .
\end{align}
\tred tensor coefficients return a \polythree object of order
$n\!=\!-2$.  \polythree objects resulting from expanding numerator
algebra in $d$ dimensions are of order $n\!=\!0$. The \polythree class
handles the arithmetic on these polynomials automatically, and the
result is another \polythree object of order $n\!=\!-2$. Since we
compute all orders at once, one may retrieve the $\epsilon^{-1}$ and
$\epsilon^{-2}$ pole coefficients, in addition to the finite value
($\epsilon^0$) without performance penalty for comparison against
other divergent pieces such as real-emission contributions, as a check
on cancellation. This also makes it simple to convert \nlox's
conventions to a different overall factor than
$S_\eps=(4\pi)^\eps/\Gamma(1-\eps)$, for instance to
$(4\pi)^\eps\Gamma(1+\eps)$ instead, without having to perform
additional recomputations at different orders in $\epsilon$ (the
finite values between the two conventions differ, and for the
conversion knowledge of the $\eps^{-2}$ pole coefficient for each PSP
is required).

\section{Benchmarking Phase-Space Points}
\label{sec:psp}

This appendix contains the collection of PSP used in the comparisons
reported in Sec.~\ref{sec:benchmarks}. They have been obtained for
center-of-mass energy $\sqrt{s}=1$~TeV and for the values of particle
masses listed in Table~\ref{tab:params}. All values are in GeV.

\begin{table}[H]
\centering
\begin{tabular}{c|c c c c}
 & E & $p_x$ & $p_y$ & $p_z$ \\ \hline
$p_1$ & 500 & 0 & 0 & 500 \\
$p_2$ & 500 & 0 & 0 & -500 \\
$p_3$ & 500.0000000000001 & 24.79419119005815 & -43.16708238242285 & -467.1321130920246 \\
$p_4$ & 500.0000000000001 & -24.79419119005815 & 43.16708238242285 & 467.1321130920246 \\
\end{tabular}
\caption{PSP used for $2\rightarrow 2$ processes with masses \{0, 0\} 
  $\rightarrow$ \{$m_t$, $m_t$\}. }
\label{tab:psp00tt}
\end{table}

\begin{table}[H]
\centering
\begin{tabular}{c|c c c c}
 & E & $p_x$ & $p_y$ & $p_z$ \\ \hline
$p_1$ & 500 & 0 & 0 & 500 \\
$p_2$ & 500 & 0 & 0 & -500 \\
$p_3$ & 500 & 26.38931223984965 & -45.94421357571205 & -497.18480813317 \\
$p_4$ & 500 & -26.38931223984965 & 45.94421357571205 & 497.18480813317 \\
\end{tabular}
\caption{PSP used for $2\rightarrow 2$ processes with masses \{0, 0\} 
$\rightarrow$ \{0, 0\}.}
\label{tab:psp0000}
\end{table}

\begin{table}[H]
\centering
\begin{tabular}{c|c c c c}
 & E & $p_x$ & $p_y$ & $p_z$ \\ \hline
$p_1$ & 500.0132299027607 & 0 & 0 & 499.9955900583434 \\
$p_2$ & 499.9955900583435 & 0 & 0 & -499.9955900583434 \\
$p_3$ & 504.1531425857718 & 26.16964177919545 & -45.56176379949805 & -493.0461320342924 \\
$p_4$ & 495.8556773753323 & -26.16964177919545 & 45.56176379949805 & 493.0461320342924 \\
\end{tabular}
\caption{PSP used $2\rightarrow 2$ processes with masses \{$m_b$, 0\} 
$\rightarrow$ \{$m_Z$, $m_b$\}.}
\label{tab:pspb0Zb}
\end{table}

\begin{table}[H]
\centering
\begin{tabular}{c|c c c c}
 & E & $p_x$ & $p_y$ & $p_z$ \\ \hline
$p_1$ & 500 & 0 & 0 & 500 \\
$p_2$ & 500 & 0 & 0 & -500 \\
$p_3$ & 410.3062535434128 & 209.5952076232996 & 319.7844730351641 & 80.83292301902506 \\
$p_4$ & 319.3952351003139 & -23.91602267647318 & -268.5728349481489 & 0.7296519976099702 \\
$p_5$ & 270.2985113562734 & -185.6791849468265 & -51.21163808701539 & -81.56257501663505 \\
\end{tabular}
\caption{PSP used for $2\rightarrow 3$ processes with masses \{0, 0\} 
$\rightarrow$ \{$m_h$, $m_t$, $m_t$\}.}
\label{tab:psp00htt}
\end{table}

\begin{table}[H]
\centering
\begin{tabular}{c|c c c c}
 & E & $p_x$ & $p_y$ & $p_z$ \\ \hline
$p_1$ & 500 & 0 & 0 & 500 \\
$p_2$ & 500 & 0 & 0 & -500 \\
$p_3$ & 403.850342409396 & 210.9997079884658 & 321.9273531813533 & 81.37458554645134 \\
$p_4$ & 323.3024476543406 & -21.68882659922996 & -273.3877478529604 & 1.940820425061027 \\
$p_5$ & 272.8472099362635 & -189.3108813892359 & -48.53960532839304 & -83.31540597151238 \\
\end{tabular}
\caption{PSP used for $2\rightarrow 3$ processes with masses \{0, 0\} 
$\rightarrow$ \{$m_Z$, $m_t$, $m_t$\}.}
\label{tab:psp00Ztt}
\end{table}

\begin{table}[H]
\centering
\begin{tabular}{c|c c c c}
 & E & $p_x$ & $p_y$ & $p_z$ \\ \hline
$p_1$ & 500 & 0 & 0 & 500 \\
$p_2$ & 500 & 0 & 0 & -500 \\
$p_3$ & 402.0189080897786 & 211.2578676637872 & 322.3212336362917 & 81.474147942932212 \\
$p_4$ & 324.4071486511459 & -20.88469923564341 & -274.7522666471455 & 2.362076641283053 \\
$p_5$ & 273.5739432590755 & -190.3731684281437 & -47.56896698914623 & -83.83622458421519 \\
\end{tabular}
\caption{PSP used for $2\rightarrow 3$ processes with masses \{0, 0\} 
$\rightarrow$ \{$m_W$, $m_t$, $m_t$\}.}
\label{tab:psp00Wtt}
\end{table}

\begin{table}[H]
\centering
\begin{tabular}{c|c c c c}
 & E & $p_x$ & $p_y$ & $p_z$ \\ \hline
$p_1$ & 500 & 0 & 0 & 500 \\
$p_2$ & 500 & 0 & 0 & -500 \\
$p_3$ & 418.7463828230725 & 283.4177855536268 & 214.8754803831553 & 221.0235731531681 \\
$p_4$ & 38.89603897017435 & -4.704398436105869 & -38.20914350134605 & 5.552642237460592 \\
$p_5$ & 263.1489010119332 & -86.76860120270908 & -121.0944656633457 & -133.2112770557442 \\
$p_6$ & 279.2086771948202 & -191.9447859148119 & -55.5718712184636 & -93.36493833488464
\end{tabular}
\caption{PSP used for $2\rightarrow 4$ processes with masses \{0, 0\} 
$\rightarrow$ \{0, 0, 0, 0\}.}
\label{tab:psp000000}
\end{table}




\bibliographystyle{elsarticle-num}
\bibliography{nlox-beta}





\end{document}